\documentclass[final,3p,times,twocolumn]{elsarticle}
\usepackage{amssymb}
\usepackage{color}
\usepackage{aas_macros}

\journal{Astroparticle Physics}

\begin{document}

\newcommand{\lsi}   {LS~I~+61~303}
\newcommand{\mo}    {M$_{\odot}$}

\begin{frontmatter}



\title{Binaries with the eyes of CTA}


\author[a]{J.M. Paredes}
%\corref{cor1}\fnref{label2}}
%\fntext[cor1]{Corresponding author.}
\ead{jmparedes@ub.edu}
\author[b]{W. Bednarek}
\author[c]{P. Bordas}
\author[d]{V. Bosch-Ramon}
\author[e]{E. De Cea del Pozo}
\author[f]{G. Dubus}
\author[g]{S. Funk}
\author[e]{D. Hadasch}
\author[h]{D. Khangulyan}
\author[i]{S. Markoff}
\author[a]{J. Mold\'on}
\author[a]{P. Munar-Adrover}
\author[j]{S. Nagataki}
\author[k]{T. Naito}
\author[l]{M. de Naurois}
\author[e]{G. Pedaletti}
\author[m,n]{O. Reimer}
\author[a]{M. Rib\'o}
\author[n,o]{A. Szostek}
\author[p]{Y. Terada}
\author[q,e]{D.F. Torres}
\author[a,r]{V. Zabalza}
\author[s]{A.A. Zdziarski}
\author[]{for the CTA Consortium}

\address[a]{Departament d'Astronomia i Meteorologia, Institut de Ci\`encies del Cosmos (ICC), Universitat de Barcelona (IEEC-UB), Mart\'{\i} i Franqu\`es 1, E-08028, Barcelona, Spain}
\address[b]{Department of Astrophysics, University of L\'od\'z, ul. Pomorska 149/153, 90-236 L\'od\'z, Poland}
\address[c]{Institut f\"ur Astronomie und Astrophysik, Universit\"at T\"ubingen, Sand 1, 72076 T\"ubingen, Germany}
\address[d]{Dublin Institute for Advanced Studies, 31 Fitzwilliam Place, Dublin 2, Ireland}
\address[e]{Institut de Ci\`encies de l'Espai (IEEC-CSIC), 08193 Bellaterra, Spain}
\address[f]{UJF-Grenoble 1/CNRS-INSU, Institut de Plan\'etologie et d'Astrophysique de Grenoble (IPAG) UMR 5274, 38041 Grenoble, France}
\address[g]{W. W. Hansen Experimental Physics Laboratory, Kavli Institute for Particle Astrophysics and Cosmology, Department of Physics and SLAC National Accelerator Laboratory, Stanford University, Stanford, CA 94305, USA}
\address[h]{Institute of Space and Astronautical Science/JAXA, 3-1-1 Yoshinodai, Chuo-ku, Sagamihara, Kanagawa 252-5210, Japan}
\address[i]{Astronomical Institute �Anton Pannekoek�, University of Amsterdam, P.O. Box 94249, 1090 GE Amsterdam, The Netherlands}
\address[j]{Yukawa Institute for Theoretical Physics, Kyoto University, Kitashirakawa Oiwake-cho, Sakyo-ku, Kyoto 606-8502, Japan}
\address[k]{Faculty of Management Information,Yamanashi Gakuin University, Sakaori 2-4-5, Kofu-si, Yamanashi 400-8575, Japan}
\address[l]{Laboratoire Leprince-Ringuet, Ecole Polytechnique, CNRS/IN2P3, F-91128 Palaiseau,
France}
\address[m]{Institut f\"ur Astro- und Teilchenphysik, Leopold-Franzens-Universit\"at Innsbruck, A-6020 Innsbruck, Austria}
\address[n]{Kavli Institute for Particle Astrophysics and Cosmology, SLAC National Accelerator Laboratory, 2575 Sand Hill Road, Menlo Park, CA 94025, USA}
\address[o]{Astronomical Observatory, Jagiellonian University, Orla 171, 30-244 Krak\'ow, Poland}
\address[p]{Graduate School of Science and Engineering, Saitama University, 255 Simo-Ohkubo, Sakura-ku, Saitama City, Saitama 338-8570, Japan}
\address[q]{ICREA, 08010 Barcelona, Spain}
\address[r]{Max-Planck-Institut f\"ur Kernphysik, P.O. Box 103980, 69029 Heidelberg, Germany}
\address[s]{Centrum Astronomiczne im. M. Kopernika, Bartycka 18, 00-716 Warszawa, Poland}

\begin{abstract}
The binary systems that have been detected in gamma rays 
have proven very useful to study high-energy processes, in
particular particle acceleration, emission and radiation reprocessing, and the dynamics of the underlying
magnetized flows. Binary systems, either detected or potential gamma-ray
emitters, can be grouped in different subclasses depending on the nature of the binary components 
or the origin of the
particle acceleration: the interaction of the winds of either a pulsar and a
massive star or two massive stars; accretion onto a compact object and jet formation; and
interaction of a relativistic outflow with the external medium. We
evaluate the potentialities of an instrument like the Cherenkov telescope array
(CTA) to study the non-thermal physics of gamma-ray binaries, which requires the
observation of high-energy phenomena at different time and spatial scales.
We analyze the capability of CTA, under different configurations, to
probe the spectral, temporal and spatial behavior of gamma-ray binaries in the
context of the known or expected physics of these sources. CTA will be
able to probe with high spectral, temporal and spatial resolution the physical
processes behind the gamma-ray emission in binaries, 
significantly increasing as well the number of known sources. 
This will allow the
derivation of information on the particle acceleration and emission
sites qualitatively better than what is currently available.
\end{abstract}


\end{frontmatter}

% \linenumbers


\section{Introduction} \label{}  
The many spectacular discoveries made in recent years by both satellite-borne ({\it
AGILE, Fermi}) and ground-based gamma-ray telescopes (H.E.S.S., MAGIC and VERITAS) have revealed a variety of new
sources of high-energy particles in the Galaxy. Among these sources we can mention star-forming regions,
accreting black holes and microquasars, early-type stars with very strong stellar winds, young isolated pulsars and
their nebulae and pulsars in binary systems. The physics of particle acceleration and interaction in the complex environment
of such astrophysical systems is extremely rich. The detection of very high energy (VHE)  gamma rays ($E >$ 100 GeV)
by the current imaging atmospheric Cherenkov Telescopes (IACT) from the systems PSR B1259$-$63 \cite{2005A&A442.1A,
2009A&A...507..389A}, LS~5039 \cite{2005Sci...309..746A}, LS~I +61 303 \cite{2006Sci...312.1771A, 2008ApJ...679.1427A} and HESS J0632$+$057
\cite{2007A&A...469L...1A}, as well as the hint of a VHE flare in the black hole binary Cygnus X-1
\cite{2007ApJ...665L..51A}, provides a clear evidence of very efficient particle acceleration in binary systems containing
compact objects (see e.g. \citep{2011arXiv1101.4843P}). In addition, one of the brightest {\em Fermi}/LAT  sources,
1FGL~J1018.6$-$5856, has been proposed to be a new gamma-ray binary \citep{2012Sci...335..189F} that could be associated with
a H.E.S.S. source \citep{2010cosp...38.2803D}. Furthermore, there are other binary systems from which VHE emission is
expected from a theoretical point of view \citep{2011MmSAI..82..182B}. Although they have not yet been detected with the
current generation of Cherenkov telescopes, their emission at HE gamma-rays ($E >$ 100 MeV) has already been reported in some
cases (e.g. Cygnus X-3, V~407 Cyg and  Eta~Carinae). It is expected that CTA will find new gamma-ray binaries, allowing
population studies that will have an impact on evolutionary models of high-mass binary systems. With a few exceptions, most of the gamma-ray binaries detected, either accreting or non-accreting sources, 
are all within 3 kpc of the Sun,  
in a volume equal to  about $\sim$~10\% of the volume of our Galaxy. Assuming a uniform
distribution, although they should follow population I stars with more objects in the spiral arms, this is consistent with $>$ 50 or
so gamma-ray binaries in our Galaxy. This number is also dependent on the duty cycle of gamma-ray emission: VHE emission in
HESS~J0632$+$057, LS~I~+61~303, PSR~B1259$-$63 is strongly dependent on orbital phase and in some sources the orbital periods can be (very) long.
With a ten times improvement in sensitivity, CTA should be able to probe for gamma-ray binaries of comparable luminosities up
to the Galactic center. CTA can thus be reasonably expected to detect a couple of dozen gamma-ray binaries. The VHE counterparts
of LS~5039,  HESS~J0632$+$057 and (possibly)
1FGL~1018.6$-$5856 were discovered in the H.E.S.S. Galactic Plane survey.  The ten times more
sensitive Galactic Plane survey planned  for CTA should thus enable many discoveries of such systems,  which are otherwise
very difficult to uncover by X-ray, optical or radio surveys.
A survey of the central
portion of the galactic plane is planned for the beginning of CTA operation (see {\citep{2012arXiv1208.5686D}), which will
pinpoint new gamma-ray  binaries candidates. 

The study of known and/or new compact binary systems at VHE is of primary
importance because their complexity allows us to probe several physical processes
that are still poorly understood. Some of these systems are extremely efficient accelerators
that could shed new light, and eventually force a revision of, particle
acceleration theory (see e.g. \cite{2008MNRAS.383..467K}). The particle injection and radiation emission mechanisms in binary systems vary
periodically due to an eccentric orbit and/or interaction geometry changes. This
may provide information on the location of the high energy particles, on the energy mechanism(s) powering relativistic
outflows, on the nature of the accelerated particles, and on the physical
conditions of the surrounding environment. The presence of strong photon fields
allows the study of photon-photon absorption and electromagnetic cascades. All these processes occur on
timescales $\lesssim 1000$ s, a proper study of which would require at least a
5$\sigma$ (standard deviations) detection for $\sim$ one hour exposure times. 

The interaction of
binary systems with the Interstellar medium (ISM) could also be powering a new
class of TeV sources, which could be resolved/detected with enough
resolution/sensitivity. For a deep study of the processes taking place in
compact binary systems we need to go beyond the present IACT's capabilities.
Below, we report on examples of numerical simulations
performed to show how the forthcoming CTA observatory \citep{2011Actis} could fulfill these
objectives.

The structure of the paper is as follows: in Sect.~\ref{stellar} we outline the
stellar gamma-ray source classes that are  idoneous targets for CTA. In
Sect.~\ref{key} we introduce key questions in high-energy astrophysics that CTA
can address and the requirements to achieve them. 
In Sect.~\ref{performance} we present the results of some
performance tests of the capability of CTA to achieve the aims.
Finally we present a summary in Sect.~\ref{summary}.

\section{Binary systems with gamma-ray emission}\label{stellar}

\subsection{\sl  Binary systems with young non-accreting pulsars}\label{plbin}

PSR~B1259$-$63 was the first variable galactic source of VHE gamma-rays
discovered \citep{2005A&A442.1A}. It has also been detected at HE by {\it AGILE} \citep{2010ATel.2772....1T} and {\it Fermi}/LAT \cite{2011ApJ...736L..10T, 2011ApJ...736L..11A}. The system contains a O9.5 Ve main sequence
donor (LS 2883) and a 47.7 ms radio pulsar orbiting its companion every 3.4
years in a very eccentric orbit (see \citep{2011ApJ...732L..10M} and references therein). Particles are accelerated in the shock between
the relativistic wind of the young non-accreting pulsar and the stellar wind of
the massive companion star \cite{1997ApJ...477..439T, dubus06, 2007MNRAS.380..320K}. 
These particles, by inverse Compton (IC) up-scattering of stellar UV photons should produce VHE gamma rays. The strong wind-wind interactions may also produce extended synchrotron radio
emission, as recently reported by \citep{2011ApJ...732L..10M}. 

The other binary systems that have been unambiguously detected at TeV energies, showing gamma-ray flux modulations
coincident with their orbital periods, are LS~5039 with $P_{\rm orb}\approx 3.9$~d  \citep{2006A&A...460..743A} and
LS~I~+61~303 with $P_{\rm orb}\approx 26.5$~d \citep{2009ApJ...693..303A}.  {\it Fermi} has also detected emission
modulated with the orbital period in both systems \cite{2009ApJ...706L..56A, 2009ApJ...701L.123A}. Although the nature
(black hole or neutron star) of the compact object in LS~I~+61~303 and LS~5039 has not yet been determined
\cite{2005MNRAS.364..899C, 2005MNRAS.360.1105C}, both systems present some similarities with PSR~B1259$-$63. They show
variable milli-arcsecond scale radio structure  \cite{2000Sci...288.2340P, 2008A&A...481...17R, 2006smqw.confE..52D}, similar
to that found in PSR~B1259$-$63. VLBA images of LS~I~+61~303  obtained during a full orbital cycle show a rotating
elongated morphology  \citep{2006smqw.confE..52D}, which may be consistent with a model based on the  interaction between the
relativistic wind of a young non-accreting pulsar and the  wind of the stellar companion \citep{dubus06}. A similar behavior
has been observed in LS~5039. This system was observed with the VLBA during five consecutive days showing an orbital
morphological variability, displaying one sided and bipolar structures, but recovering the same morphology when observing at
the same orbital phase \citep{2012arXiv1209.6073M}. The broadband emission from radio to VHE gamma-rays of the
three sources is variable and periodic, peaking at MeV-GeV energies.

For LS~I~+61~303 and LS~5039, the GeV and TeV emission are well anticorrelated. In particular, in LS~5039, the GeV
emission peaking around the compact object superior conjunction/periastron, and the TeV radiation around inferior
conjunction/apastron. The GeV and TeV spectra are also roughly anticorrelated, with the GeV emission getting harder for lower
fluxes, and the TeV emission for higher ones (e.g., \cite{2006A&A...460..743A, 2012ApJ...749...54H}).  In both sources, the behavior is more
or less compatible with radiation produced by IC and moderate gamma-ray absorption, processes through which the changing geometry 
along the orbit induces a modulation in both flux at TeV and GeV, and  spectrum at TeV (and GeV if cascades were important).
Additional effects like varying radiative and adiabatic losses could also affect spectra  and fluxes \citep{takahashi09}. We note that 
also  IC e$^\pm$ pair moderate cascading can be important \cite{1997A&A...322..523B, 2000A&A...362..646B, 2006MNRAS.368..579B}. However, the relatively low flux around 10~GeV in both LS 5039 and LS I +61 303, below the extrapolation of data and model
predictions, strongly indicates that the emitter, although of likely leptonic+IC nature, should be quite complex (and probably
located in the periphery of the binary system for LS 5039).

The most recent addition to the selected group of gamma-ray binaries emitting up to very high energies is HESS~J0632$+$057
\cite{2009ApJ...690L.101H}. The source was initially detected by the H.E.S.S. experiment \citep{2007A&A...469L...1A},
but the subsequent non detection by VERITAS excluded it as a steady gamma-ray emitter \citep{2009ApJ...698L..94A}. The
gamma-ray variability was confirmed recently by VERITAS and MAGIC which reported an increase of gamma-ray flux during 2011
February 7--9 (see Refs.~\citep{2011ATel.3153....1O} and \citep{2012arXiv1203.2867T}, respectively). The increase in TeV
flux coincides with the time   of a large X-ray peak, that could imply that the same population of electrons is producing 
the X-ray and TeV emission \citep{2011ApJ...737L..11B},   similarly to that found in LS~I~+61~303 \citep{2009ApJ...706L..27A}
and LS~5039 \citep{takahashi09}.  The source shows an X-ray periodicity of $321\pm5$~d which has been associated to its
orbital period \citep{2011ApJ...737L..11B, 2012MNRAS.421.1103C}. No X-ray pulsations have been detected so far
\cite{2011ApJ...737L..12R}. In the radio band, VLBI observations reveal an extended and variable non-thermal radio source
\citep{2011A&A...533L...7M}, with its position being compatible with the B0pe star MWC~148. Overall, these results are
very similar to the multi-wavelength data obtained for the other gamma-ray binaries, although HESS~J0632$+$057 displays a
spectral energy distribution (SED) one order of magnitude fainter.  This may provide new constraints on the luminosity
distribution of this subclass of binary systems in our Galaxy.

A promising candidate to be a new VHE gamma-ray source associated to a binary system has emerged recently. 1FGL~J1018.6$-$5856 was discovered by searching for
periodicities of {\it Fermi} sources, and shows intensity and spectral
modulation at GeV energies with a 16.6-day period
\citep{2012Sci...335..189F}. Given the variability of the proposed
X-ray and radio counterparts, and the spatial coincidence with an
O6V((f)) star, these authors proposed this source as a new gamma-ray
binary.

\subsection{\sl  Microquasars}

Microquasars (MQ) are X-ray binaries (XRB) with relativistic jets emitting non-thermal radio emission through synchrotron
radiation. The gravitational energy released by accretion feeds the relativistic radio jets and powers the non-thermal
emission. Two MQs, Cygnus X-3 \cite{2009Natur.462..620T, 2009Sci...326.1512F} and Cygnus X-1 \citep{2010ApJ...712L..10S}, have been
recently detected at HE gamma-rays and there is evidence of VHE gamma-ray emission associated with Cygnus X-1 during a flare
\citep{cygx1} but not with Cygnus X-3 \citep{2010ApJ...721..843A}.  The gamma-ray emission detected by {\it
Fermi} from Cygnus X-3 is modulated with the orbital period of the system and is correlated with the radio emission, which is associated with the relativistic jets.
Other MQs, like SS 433
\citep{2011arXiv1110.1581Z}, GRS 1915$-$105 \citep{2009A&A...508.1135H} or Sco X-1 \citep{sco11}, have also been observed at
HE and VHE but they have not been detected yet. They show however synchrotron radio emission that probes their capability to
accelerate particles up to relativistic energies, and their transient nature makes necessary to look at the sources at the
right time to detect their high-energy radiation \citep{vila10}.  The presence of strong photon
fields provided by the companion star may offer a good scenario for the production of gamma-rays through IC 
processes \citep{2002A&A...385L..10K}. Furthermore, proton-proton interactions could also take place in the collision between
relativistic protons and the stellar wind \citep{2003A&A...410L...1R}. However, photon-photon absorption may also take place
in the inner regions of the system, and the  VHE photons may be strongly attenuated  \citep{1993MNRAS.260..681M, 1997A&A...322..523B, 2006A&A...451....9D}. 

It is noteworthy that, potentially, the non-thermal processes leading to radio and high-energy
radiation in high-mass microquasars can be difficult to distinguish from those in young non-accreting pulsars. As pointed
out, e.g. in  \cite{2010A&A...512L...4P} and \cite{2011PASJ...63.1023B}, 
the complex
and variable/periodic radio structures found in gamma-ray binaries, some with a compact object of unknown nature, may be also produced by jet
disruption or gamma-ray absorption and secondary radio emission, respectively. In addition, the processes underlying or
affecting the high-energy radiation, for instance in LS~5039 and LS~I~+61~303 (as described in Sect.~\ref{plbin}), would also
take place in high-mass microquasars. Therefore, despite some differences in the non-thermal radiation are expected, they are
likely to be more of quantitative than of qualitative nature, and only high-quality data may be able to allow the
identification of the underlying engine.

\subsection{\sl Collision of the outflow with the interstellar medium}

 The termination regions of MQ jets can also generate non-thermal emission. The detection of VHE emission from those regions would represent a new type of gamma-ray emitter. In a scenario similar to that found in the Sedov-expansion phase of supernova remnants (SNRs), MQ jets propagating into the medium are eventually decelerated, developing forward/reverse shocks into the ISM/jet-ejection. In these shocks particles can undergo efficient acceleration and could produce gamma-ray emission through IC, relativistic Bremsstrahlung and proton-proton collision 
%$\pi^{0}$-decay 
processes. The dynamics of these interactions (see e.g. \citep{2002A&A...388L..40H}) and the possibility of
the production of gamma-ray emission (\citep{2009A&A...497..325B}, \citep{2011MNRAS.410..978Z}) have been recently addressed. The expected fluxes may depend on the MQ jet power, the source age, the external medium particle density profile, and the duty cycle if transient, and could reach values $F_{\rm E>10GeV}\sim$~few~$\times 10^{-14}$~erg~cm$^{-2}$~s$^{-1}$, which is a few times below the sensitivity level of the current IACTs. These sources may be therefore good targets for the improved capabilities of CTA.

Large-scale interactions are also expected in the case of binaries hosting
a young pulsar, when mixed stellar and pulsar winds interact with the
environment. The expected radiative outcome of this interaction is similar to
that of microquasar jets \citep{2011A&A...535A..20B}.

\subsection{\sl Colliding wind binaries}
%\smallskip\\

 Hot stars can generate strong winds and form colliding wind binary systems (CWB). 
 %that may significantly affect the environment. 
 Shocks are expected to form in massive star binaries, in the region where the winds from both stars collide. Non-thermal synchrotron emission from the colliding wind region in one source has been detected \citep{2005ApJ...623..447D}, which indicates the presence of highly relativistic electrons (see also \citep{sugawara10}). These systems may also be embedded in dense photon fields where IC losses would be unavoidable, making CWBs potential high-energy emitters \cite{1993ApJ...402..271E, 2001A&A...366..605B}.  
 
 An extreme example is the Eta Carinae system \citep{2008MNRAS.384.1649D}. Gamma-ray emission has been theoretically predicted from this source (see e.g. \citep{2006ApJ...644.1118R, 2011A&A...530A..49B}) and the emission has been tentatively confirmed recently by the {\em Fermi}/LAT \citep{2010ApJ...723..649A, 2012arXiv1203.4939R} and {\it AGILE} \citep{2009ApJ...698L.142T} instruments. The predominant GeV emission of Eta Carinae, shown in the top panel of Fig.~\ref{carinae}, seems to agree with what is expected from IC and/or neutral pion decay processes in such type of system. At VHE, Eta Carinae has not been detected so far \citep{2012MNRAS.424..128H}. The reported HE flux levels and the spectrum make however this source a good target for CTA, since it will provide a significantly improved sensitivity at energies in the range 30 to 100~GeV as compared to present IACTs. MAGIC observations of WR~146 and WR~147 produced the first bounds on the high-energy emission from Wolf-Rayet binary systems \cite{2008ApJ...685L..71A}.

\begin{figure}[]
      \includegraphics[width=0.48\textwidth,angle=0]{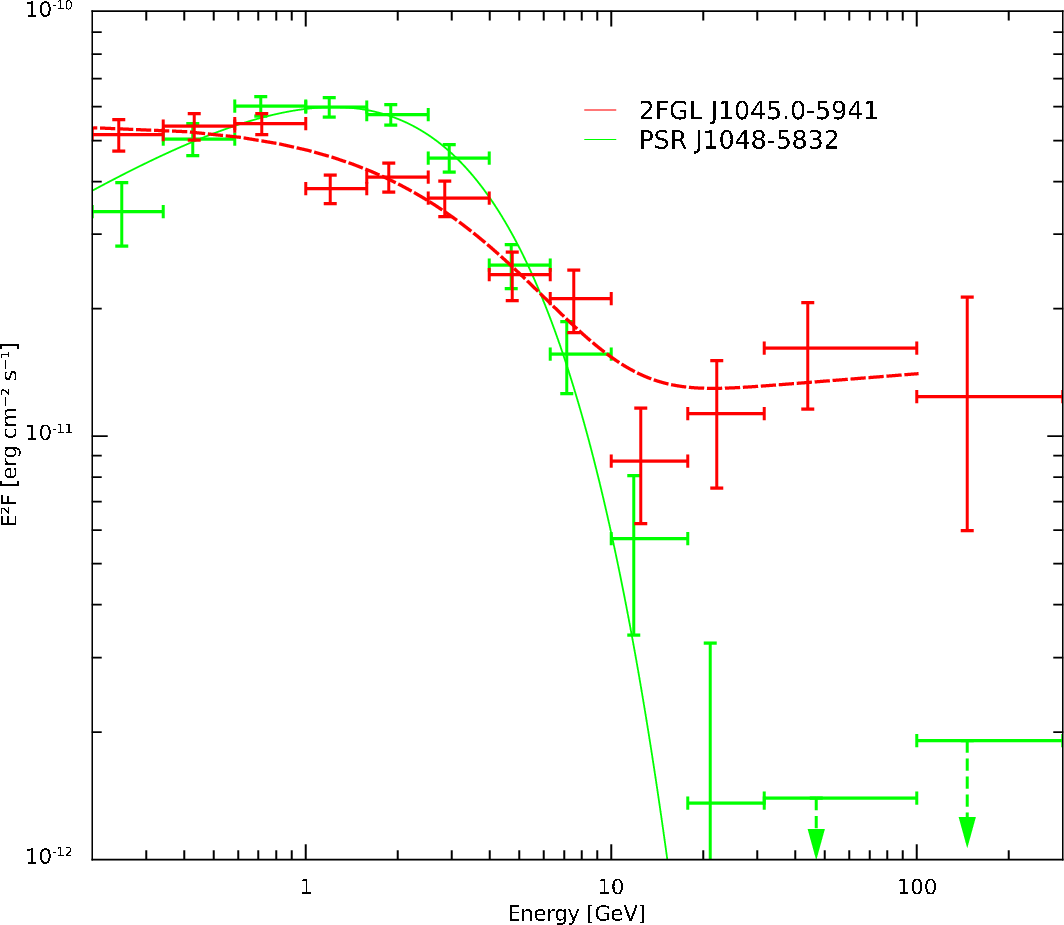}
      \includegraphics[width=0.46\textwidth,angle=0]{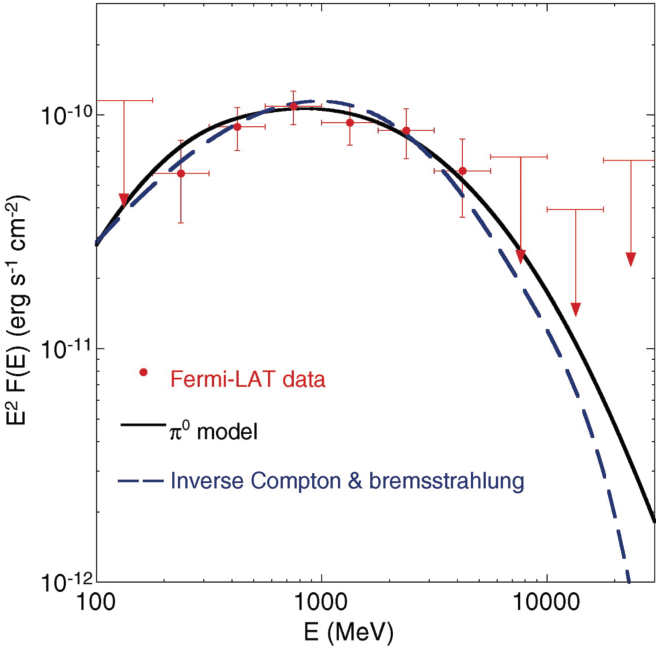}
      \caption{{\it Top}: Spectral energy distributions (SED) of two  \emph{Fermi}/LAT
      sources at the position of Eta Carinae: 2FGL J1045.0$-$5941
      (red points) and PSR J1048$-$5832 (green points). The
      best-fit model for the average spectrum is shown as a red dashed line
      (\citep{2012arXiv1203.4939R}). {\it Bottom}: SED of V407 Cyg measured over
      the period from 10 March  to 29 March 2010.
      \label{carinae}}
\end{figure}

\subsection{\sl Cataclysmic binaries}

 Regarding cataclysmic binary systems, variable high-energy gamma-ray emission has been recently detected from an optical nova of the 
symbiotic star V407 Cygni \citep{2010Sci...329..817A}. According to the source SED shown in the bottom panel of
Fig.~\ref{carinae}, the extrapolated VHE gamma-ray fluxes are too low for current IACTs sensitivities, and even with its improved capabilities, CTA may require of relatively long exposure times to detect it. 

\section{Key questions and CTA requirements}\label{key}

\subsection{Relativistic outflows in binary systems with young non-accreting pulsars}

The rapid rotation of the magnetosphere of PSR~B1259$-$63 generates a highly
relativistic wind that spins down the young neutron star. The properties of
such winds are a long-standing issue of astrophysics. Constraints on pulsar
winds have been inferred from observations of their interaction with the
surrounding medium: the supernova ejecta in pulsar wind nebulae (PWN) or the ISM
in isolated, fast moving pulsars. In a binary system, the pulsar wind interacts
with the stellar wind of its companion in a periodic fashion. The binary nature
determines the matter and radiation densities to be sampled by the pulsar during its
orbital motion, and allows the production of relativistic outflows in the asymmetric postshock region of the pulsar wind. Phase-resolved spectra of binaries powered by a young pulsar 
offer novel tests of pulsar wind models on spatial scales $\sim10^4$--$10^5$
times smaller than in isolated objects.  
 
The study of pulsar winds requires phase-resolved light curves and spectra from binary
pulsars. The systems LS~5039 and LS~I~+61~303 have been suggested to be powered
by a young pulsar \citep{dubus06}.  The orbital periods of these systems range from a few days (LS~5039) to years (PSR~B1259$-$63). In LS~5039 the spectrum changes drastically between orbital
phases of high flux as compared to phases of low flux. PSR~B1259$-$63 and
LS~I~+61~303, on the other hand, are only detected during high flux phases.
Information on the low flux phases is therefore essential for modeling purposes,
and this requires the increased sensitivity of CTA. An extended spectral range
is also crucial to study the connection to the MeV-GeV emission (due e.g. to IC
emission from cascade electrons) and to constrain the high-energy cut-offs
linked to particle acceleration. 

Gamma-ray binaries, in particular LS~5039, present a very high acceleration
efficiency \cite{2008MNRAS.383..467K, takahashi09}.  CTA good
sensitivity in the range 10--100 TeV will allow us to probe the acceleration
mechanism, the strength of the magnetic field, intrinsic orbital/short timescale
variability of the accelerator/emitter, and the emitting particle nature. This
will be possible since the opacity and orbital dependence of photon-photon
absorption, and the angular IC dependence, sources of extrinsic variability, are
smaller in this energy range. In addition, the spectrum of the emission is
expected to be different depending on whether the emitting particles are protons
or electrons.

Finally, several studies have tried to synthesize the population
of high-mass X-ray binaries from assumptions on the progenitor distributions and their evolutionary paths. Some include
predictions on the number of binaries in the spinning down pulsar+massive star stage as a byproduct \cite{meurs89,
portegies96, portegies98}. They are quite uncertain, with numbers ranging from 100 to 1000 depending on the details of
binary evolution and on the adopted lifetime of this phase.  
Constraining the population of PSR~B1259$-$63-like objects is
important as this impacts HMXB populations but, ultimately, also the population of double neutron stars
that are targeted by Virgo/LIGO. 

\subsection{The accretion/ejection link in microquasars}

The possible detection of Cygnus~X-1 by MAGIC may indicate that MQs are able to
emit VHE gamma-rays. These systems, like active galactic nuclei (AGN) and gamma-ray bursts (GRBs), are powered by accretion.
How and where the energy is released are still poorly understood. Part of the
material is ejected at relativistic speeds and this is invariably associated
with non-thermal emission. Importantly, emission from both accretion and
ejection processes is detected in MQs. Their interplay can be studied
when the mass accretion rate changes during outbursts. The MQ environment is
better constrained and the timescales are more accessible than in GRBs and AGNs.
Emission beyond a few MeV in some MQs found by \textit{CGRO/COMPTEL} is essentially {\it terra incognita},
probably because of variability and lack of sensitivity. Exploring the VHE domain and connecting it to the
known anti-correlations between thermal X-ray emission (from accreting material)
and non-thermal radio emission (from the jet) should bring new insights into
the accretion/ejection processes. Determining the time variability of the TeV to
X-ray spectrum will help to know whether the jets have electron-positron or
hadronic components, and will help as a consequence to understand the physics of the jet
production. Interestingly, powerful radio flares are likely associated with the production of gamma 
rays \cite{2009Sci...326.1512F, 1999MNRAS.302..253A, 2012MNRAS.421.2947C}. As these radio flares occur in a specific area of the hardness intensity diagram,  relatively precise predictions of a suitable trigger for CTA can be made.

Achieving this goal requires the monitoring of selected sources or
target-of-opportunity (ToO) quick response with enough sensitivity to detect flaring episodes
on $\sim$~hour timescales. The discrete ejections observed in X-rays, infrared and radio
in GRS~1915+105, for instance, occur on such timescales, and these could also be detected in gamma rays. 
The full CTA array should detect a 10-mCrab flare in one hour,
corresponding to a luminosity above 100 GeV of about $10^{-5} L_{\rm Edd}$ for a
10 M$_{\odot}$ black hole at the Galactic center distance. A significant advantage of CTA compared to current IACTs is subarray observations of specific sources. This will allow long-term
monitoring that is currently too time-consuming for presently operating IACTs. A 0.1-Crab flare would be
detected in one hour by a HESS-like subarray and could trigger follow-up
observations of the whole array. Relating this emission to the known
disk/jet/corona phenomenology will require on the one hand multi-wavelength
campaigns, and on the other hand to have the possibility of a quick CTA response
to ToO's from {\em Fermi}/LAT, X-ray satellites or radio interferometers like LOFAR and the SKA pathfinders MeerKAT, ASKAP and ATA. We note that ASKAP and LOFAR are starting all-sky transient monitoring in 2012.

\subsection{Collision of the outflow with the interstellar medium}  

 MQs are capable to transfer large amounts ($\sim10^{49}-10^{51}$~erg) of kinetic energy and momentum to the
surroundings through discrete or continuous ejections. Significant non-thermal gamma-ray emission in the jet/medium shocked
regions could be produced, although dense ambient conditions are required. ISM particle densities can vary from
$\sim$10$^{-3}$ cm$^{-3}$ (hot ISM regions) to $\sim10^{4}$~cm$^{-3}$ (inside molecular clouds). In the case of typical
galactic densities $\sim 1$~cm$^{-3}$, the predicted VHE fluxes above 100 GeV are a few $\times
10^{-14}~$erg~s$^{-1}$~cm$^{-2}$, accounting for both relativistic Bremsstrahlung and p--p interactions. The size of the
interaction structures would be $\sim$1~pc, i.e. $\sim$1~arcminute for a source located at 3 kpc. The detection of VHE
emission from the jet/medium interaction structures would reveal a new type of VHE emitter, probing the
acceleration/radiation from large-scale shocks and constraining fundamental properties of galactic jets (e.g. energetics,
magnetic field, etc.).

To detect the jet/medium interaction region requires to improve the present IACTs
sensitivity ($\sim$few$\times10^{-13}$ TeV s$^{-1}$ cm$^{-2}$) about one order of
magnitude and to have an angular resolution of 1 arc-minute. Additionally, the possibility of probing the $\lesssim$~50~GeV energy range would be extremely helpful, since this is where most of the gamma-ray flux is expected to be produced. There are no
required special timing/monitoring CTA configurations, since the emission is expected
to be steady and the source injection variability is supposed to be smoothed out
at the large termination regions. 

Similar requirements to those of jet termination regions are
met also when observing the termination of outflows produced in pulsar
binaries.

\subsection{Colliding winds of massive stars in binary systems} 

 VHE emission is theoretically expected from binary systems with high-mass loss and high-velocity winds. These systems display
some of the strongest sustained winds among Galactic objects and have the
highest known mass-loss rate of any stellar type. Colliding winds of massive
star binary systems are potential VHE gamma-ray emitters, via leptonic and/or hadronic processes after acceleration of
primary particles in the collision shocks. 

The detection of VHE emission from colliding winds
requires an improved sensitivity with respect to current IACTs. To further study these
systems in case of detection, phase-resolved light curves and
spectra would be required. Such spectra could give us a clue to understanding the physical processes behind the emission, since the
non-thermal particle distribution strongly depends on the shock conditions at each orbital phase.
A low-energy array is favoured; a cutoff at a level of $\sim$100 GeV due to the modest shock velocity and 
finite size of the acceleration zone is predicted, and CTA should be able to
operate at a lower energy threshold than present IACTs. Finally, the orbital
distances between the two stars ($\sim  10^{13} - 10^{15}$ cm) make the
emission region too small to be resolved out by the current designs of CTA
configurations. Angular resolution is therefore not a requirement in this case. 

\section{Performance tests}\label{performance}

Several of the studies herein presented are meant as examples of the capability of CTA in comparison with current IACTs;
thus we take here several of the already detected gamma-ray binaries, for which a well known set of assumptions can be
established. We have used Monte Carlo simulations of the sensitivity of the array for several possible configurations in
order to explore the capability of CTA to study  binary systems. In particular, we have conducted simulations using configurations \texttt{B}, \texttt{D}, \texttt{E}, \texttt{I}, \texttt{NA} and \texttt{NB}, and subarrays \footnote[1]{The first number indicates the number of telescopes, the second indicates the type (1 for small size telescope, 2 for medium and 3 for large), and the last number indicates the separation in meters. The subarray \texttt{s9-2-120} is composed of nine medium sized telescopes located at a distance of 120m from each other.} \texttt{s4-2-120}, and \texttt{s9-2-120}. The southern configurations are optimized for low energies (\texttt{B}), high energies (\texttt{D}), balanced with a focus on low energies (\texttt{E}), and balanced with a focus on medium and high energies (\texttt{I}). The northern configurations are optimized for low energies (\texttt{NA}), and balanced (\texttt{NB}). The subarrays are HESS-like arrays (\texttt{s4-2-120}) and expanded HESS-like arrays (\texttt{s9-2-120}). All details can be found in Hinton et al., this issue.

\subsection{CTA flux error reduction in known TeV sources}\label{error}

Accurate estimates of the flux, spectral shape, and evolution of known TeV sources are
very important for constraining the physical parameters of the high-energy
emitting region. This is even more needed when there are several parameters that
have to be left free when fitting data. To explore the CTA capability to derive
observables and constrain theoretical models, we simulated the CTA response on LS\,5039. This source
might not be representative of the class of binaries, but will allow us to 
compare the improvement from CTA data on the present generation of IACTs.
We based our simulations on the results obtained by H.E.S.S. on the
source, simulating the CTA response under similar conditions (above 1
TeV). Since the H.E.S.S. data
were taken over a long time span and under different zenith angles, the energy
threshold was not constant. 
To make a fair comparison with the H.E.S.S. analysis, we treated our simulated data the same way as the H.E.S.S. collaboration did  \citep{2006A&A...460..743A}. The simulated counts and the flux normalization were extracted above 1\,TeV assuming an
average photon index derived from all data: $\Gamma = 2.23$ for
$\mathrm{d}N/\mathrm{d}E\sim E^{-\Gamma}$. 
Based on the H.E.S.S. results we assumed a sinusoidal shape of the light curve 
with a period of 3.9 d. For each phase point, and using configuration \texttt{I}, 
we then evaluated the light curve and simulated spectra of the form described above 
as seen by CTA, for a certain observation
time in each phasogram bin.
By integrating these spectra, we got the flux above
100\,GeV in each bin.  We then extrapolated the obtained flux value to the integral
flux above 1\,TeV and propagated the error correspondingly. The result for 70\,hours
of exposure time of CTA is shown in 
Figure~\ref{hess_cta_same}.  The improvement of observations by CTA is clearly
visible. The error bars are reduced by a factor of $\sim$2--4 with respect to
the H.E.S.S. data points, assuming the average photon index of $\Gamma = $2.23.
This can be taken as a direct comparison to the published H.E.S.S. results.
However, the spectral index of the LS\,5039 VHE emission changes as a function
of orbital phase. A variation of the spectral index affects the error
estimate. Softening the assumed photon index ($\Gamma\sim$3), CTA data
would improve the errors by up to a factor of 7, whereas a harder
spectrum ($\Gamma\sim$2) results in a minor error reduction of less than a
factor of 2. We have also performed
similar simulations but taking a time binning of 10, 20, 50 and 100 bins per full phase period and using a binning of 14 and 28 minutes.  
With 10 bins per full orbital period the sinusoidal shape can hardly be seen,
whereas with 20 bins a sine function can be fitted to the data points.  Taking 50 bins
of 28 minutes even substructures can be resolved. To obtain the same results
and similar error bars like those reported by H.E.S.S., CTA would only need $\sim$1/6th of its observation
time, that is, 50 bins of 14 min each, pointing towards the possibility of performing a long-term
monitoring of the global behavior of the source, and accessing the
duty cycle of the observed features, if any.

Furthermore we studied the minimum observation time for CTA to detect the period of LS\,5039 in comparison with the H.E.S.S.
one. To do so, we simulated CTA observations using a sine function over time that reproduces the time structure of the
H.E.S.S. flux points. From the simulated CTA observations we derived flux points for each time bin and used those to
construct the power spectrum of LS\,5039. Whereas H.E.S.S. used 160 bins of 28 minutes ($\sim$70 hours in total) to detect
the 3.9 days  period of the system, CTA could detect the period with more than 5 $\sigma$ with only 160 bins of 3 min (8
hours in total). This would be a significant reduction of observation time for CTA. It has to be kept in
mind that the significance of the period estimation in the H.E.S.S. data is
larger than 5 $\sigma$ (i.e. 8 $\sigma$), as all the data available at the time
were used.

We studied the modulation of the photon index and the flux normalisation with the orbital period for a source like 
LS\,5039. To compare with the H.E.S.S. measurements, we
assumed 7~hours of observation time for each phase bin and simulated the CTA spectra
for each phase bin with the spectral parameters obtained by H.E.S.S.  (photon
index and differential flux at 1\,TeV). By fitting these simulated spectra,
we obtained the fit parameters with the corresponding error. The results are
shown in Figure~\ref{ls_index_n0_simulations}. The direct comparison of the
errors of the H.E.S.S. and CTA measurements shows that observations with CTA
can reduce the errors on the spectral parameters by a factor between 2 and
4.5.

   \begin{figure}
  % \centering
   \includegraphics*[width=0.51\textwidth,angle=0,clip]{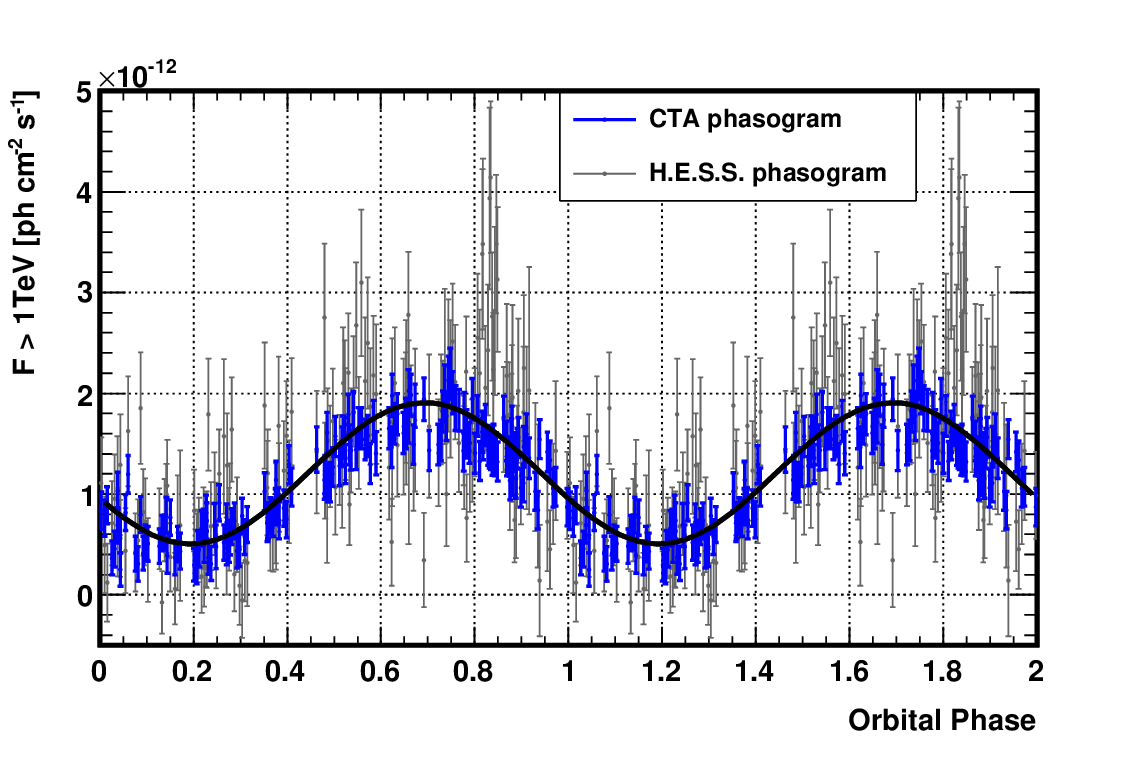}
   \caption{\label{hess_cta_same}CTA simulations of LS\,5039. Phasograms of H.E.S.S. (gray) and simulated CTA observations (blue). With CTA observations the error on the flux can be reduced by a factor 2 -- 4 above 1\,TeV. }
    \end{figure}

   \begin{figure}
   \centering
   \includegraphics*[width=1.0\linewidth,angle=0,clip]{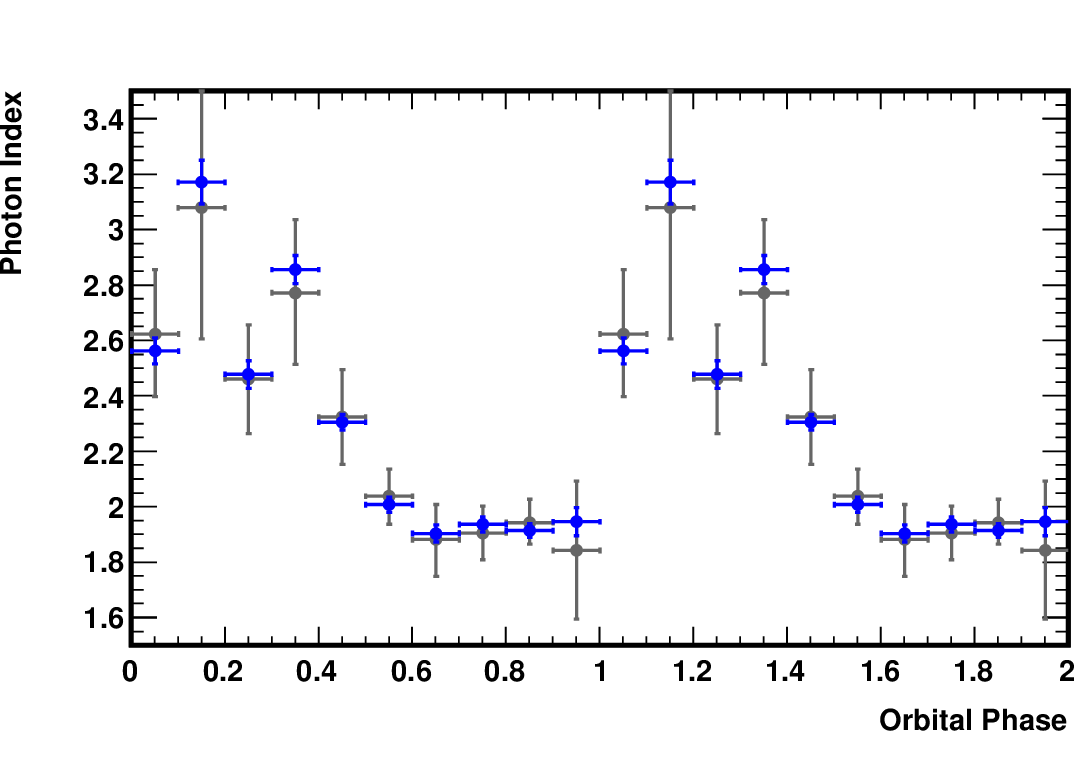}
   \includegraphics*[width=1.0\linewidth,angle=0,clip]{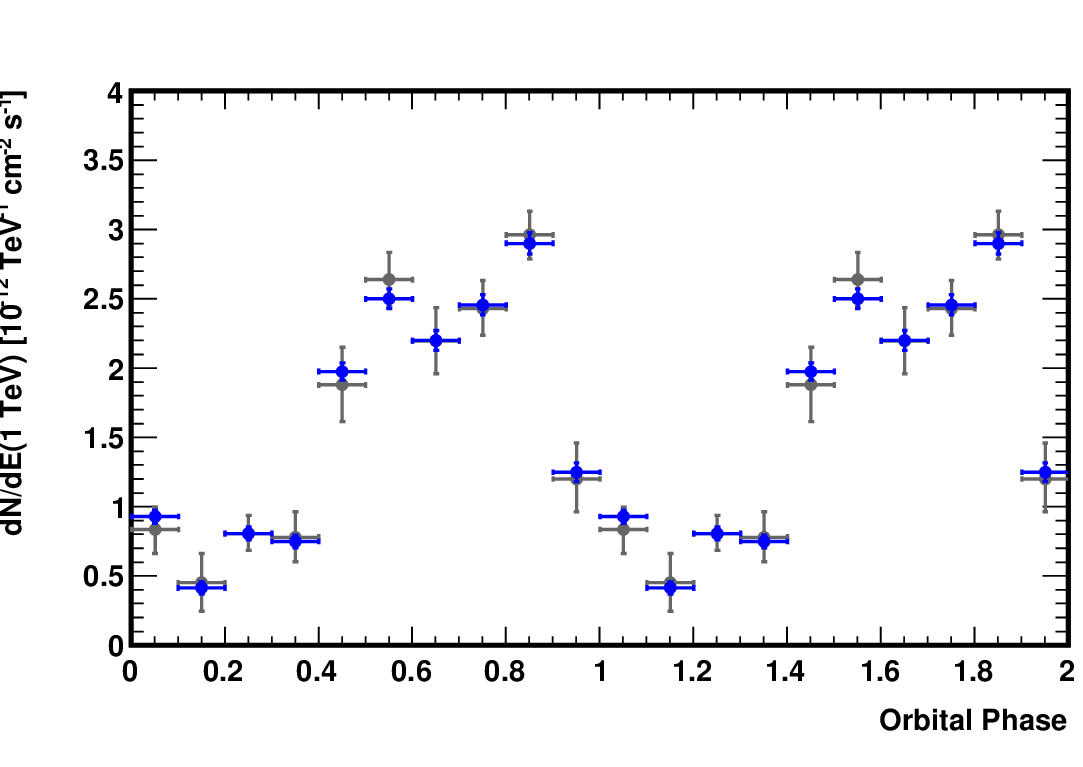}
   \caption{\label{ls_index_n0_simulations}CTA simulations of LS\,5039. 
   \textit{Top: }Photon index versus phase (CTA blue, H.E.S.S. gray). 
   %\textit{Bottom left: }Ratio of the erros of the  H.E.S.S. and CTA photon index measurements. 
   \textit{Bottom: }Flux normalisation versus phase (CTA blue, H.E.S.S. gray). 
   %\textit{Bottom right: }Ratio of the errors of the H.E.S.S. and CTA flux normalisation measurements.}
   }
    \end{figure}

The larger sensitivity of CTA would allow tracking the behavior of a source in shorter timescales. In particular, it would allow comparing with predictions of the spectral evolution of a source such as LS 5039, even at the minimum of its TeV flux.
As an example, we used the spectra in
phases 0.2 and 0.3 as derived by \citep{2008ApJ...674L..89S}, where electromagnetic cascades were 
included.
In Figure~\ref{phase_spectra}  we show the results of our simulations: in the top panel, the two simulated spectra are plotted, assuming an observation time of 5\,hours.
Since the reconstruction of the energy spectra in true energy requires a complicated unfolding
procedure we conservatively choose to compare the two spectra on the level of the excess events as
a function of reconstructed energy. The two corresponding distributions
are shown in the middle panel of Figure~\ref{phase_spectra}.
The distributions are compared to each other by calculating the residuals between the two,
which is shown in the bottom panel.
%They are clearly not compatible with 1 which one would expect if the two spectra were similar.
The probability of these spectra to be consistent (i.e. to originate from 
the same original distribution) is $\ll$1\%.
We conclude that CTA would easily distinguish between spectra at different phase bins.

   \begin{figure}
  % \centering
   \includegraphics*[width=1.0\linewidth,angle=0,clip]{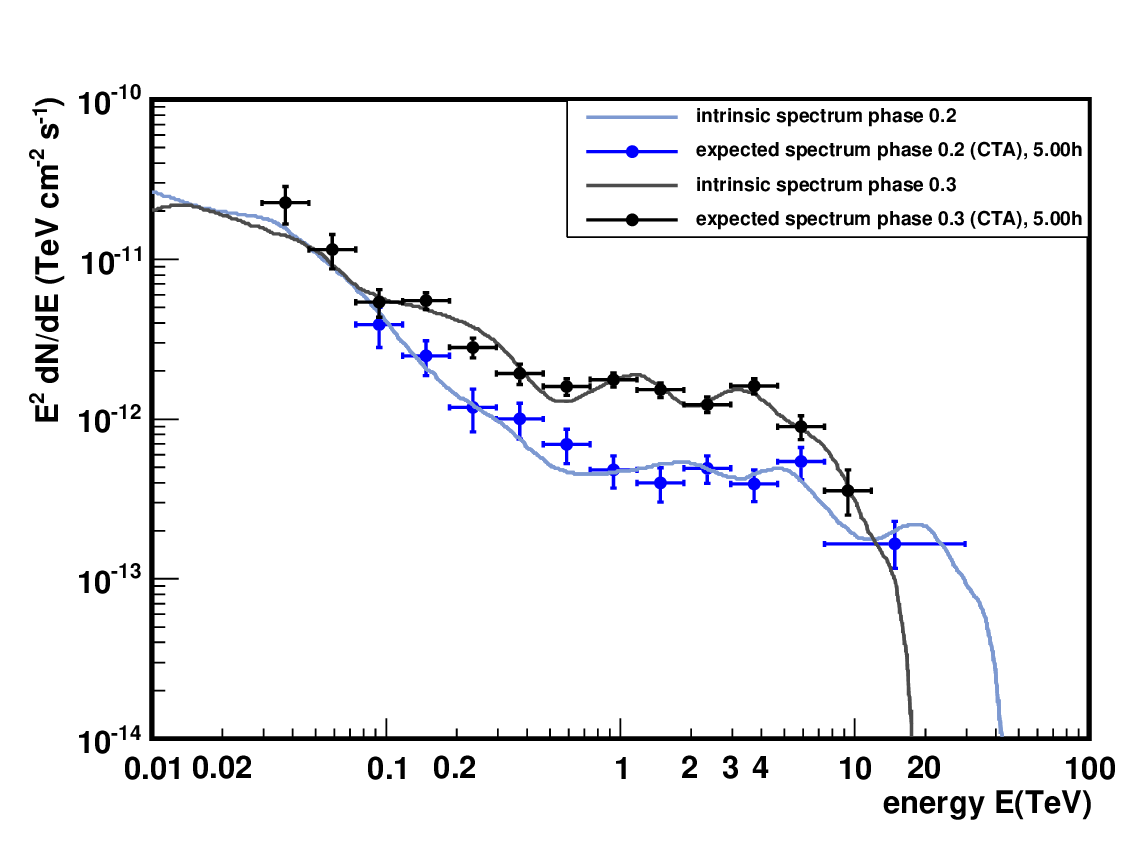}
   \includegraphics*[width=1.0\linewidth,angle=0,clip]{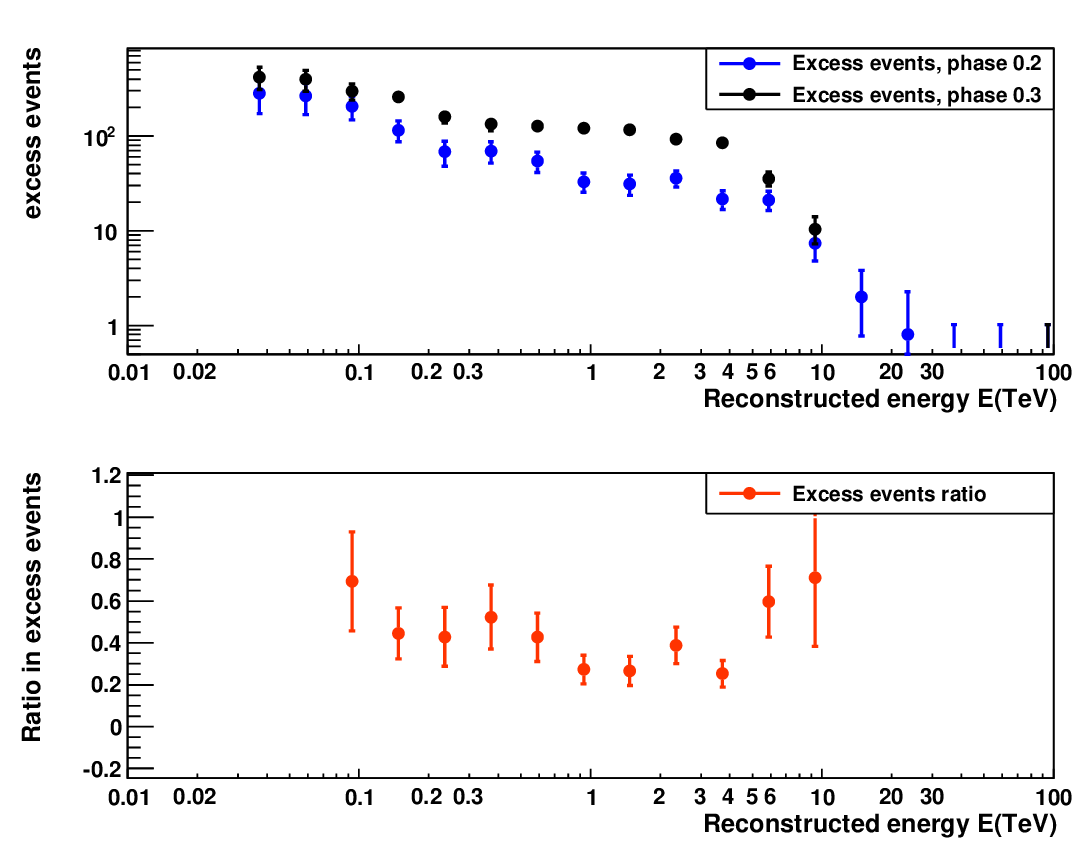}
   \caption{\label{phase_spectra} CTA simulations of LS\,5039. 
   \textit{Top: }Intrinsic (line) and simulated (data points) energy 
spectra for phases 0.2 (blue) and 0.3 (black). The assumed integration time is 5\,h. 
   \textit{Middle: }Excess events obtained from the simulations above versus reconstructed energy. 
   \textit{Bottom: }Residuals of the excess events distributions from the two distributions above.
   In all three panels a very clear difference between the two spectra can be seen.}
    \end{figure}

\subsection{Short timescale flux variability}\label{fast}

We have studied the capabilities of CTA to detect short timescale flux and spectral
variabilities from gamma-ray binaries. The perfect candidate for such studies
is \lsi. A simultaneous multiwavelength campaign on this source resulted in
the discovery of correlated X-ray/VHE emission with the MAGIC IACT and
the X-ray satellites \emph{XMM-Newton} and \emph{Swift} in the energy range 0.3--10 keV \citep{2009ApJ...706L..27A} (see however \cite{2011ApJÉ738É3A, 2009ApJ...700.1034A}). Additionally, fast X-ray
variability on timescales of a thousand seconds has been detected
\citep{2010MNRAS.405.2206R, 2006A&A...459..901S}, so it is natural to think that
such fast variability may be present in its VHE emission. The detection of such a variability would have
strong implications on the location and size of the non-thermal emitter. Several
theoretical models put forward until now consider the possibility of short
timescale ($\sim$100\,s), correlated X-ray/VHE variability as the result of
leptonic emission in small scale instabilities and shocks within, e.g. a
relativistic jet \cite{2006A&A...447..263B} or the interaction of the pulsar wind with
a clumpy stellar wind \cite{2010MNRAS.403.1873Z}. The detection of such a feature with
CTA will provide information on the dynamical mechanisms underlying the VHE
emission, which is a crucial step in understanding the apparently erratic
variability of \lsi. This knowledge will be also valuable in understanding other
gamma-ray binaries with complex stellar wind geometries owing to the decretion
disk of Be stars, such as PSR~B1259$-$63 and HESS~J0632$+$057.

Under the assumption that the X-ray/VHE emission correlation found by
\cite{2009ApJ...706L..27A} holds at shorter timescales, we took the X-ray
light curve from a $\sim$100\,ks long \emph{Chandra} observation
\cite{2010MNRAS.405.2206R} and computed the expected VHE emission from the
source. We considered the large statistical errors in the correlation by varying
the parameters following normal distributions with a width equivalent to the
uncertainties for each of the points in the light curve.

\begin{figure*}
    \includegraphics[width=0.50\linewidth,angle=0,clip]{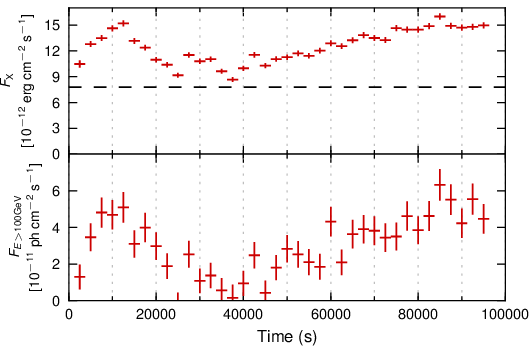}
    \includegraphics[width=0.50\linewidth,angle=0,clip]{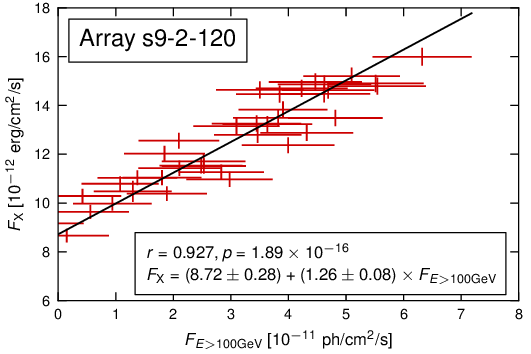}
    \caption{CTA simulations of  \lsi\  during 100\,ks, a time span similar to that of a 10 night observation
  campaign with around three hour of observation per night.
 \textit{Left:} X-ray and VHE light curves
        for CTA \texttt{s9-2-120}. Measured X-ray flux (top, 0.3--10\,keV)
%        in units of $\mathrm{erg\,cm^{-2}\,s^{-1}}$
        and simulated VHE fluxes
        obtained with the subarray \texttt{s9-2-120} (bottom, $E> 100$\,GeV in
        units of $\mathrm{ph\,cm^{-2}\,s^{-1}}$) for the case of a campaign of
        ten nights of observation and binning time of 2500\,s.  \textit{Right:}
        X-ray/VHE correlation for CTA \texttt{s9-2-120}. X-ray/VHE correlation
        plot from the ten-night, subarray \texttt{s9-2-120} observation plotted
        in 2500\,s bins.  The Pearson correlation factor, the chance probability
        of correlation and the parameters of the fitted correlation are shown as
        insets.\label{corrsub}}
\end{figure*}

We considered two feasible observation scenarios. In the first, the full CTA
\texttt{I} array is used to perform a deep, short observation of \lsi.  In this
case, the observation length would be of the order of 4\,h, which is the typical
time span for which a source is visible under optimum conditions for a single
night. To simulate this scenario, we selected 10\,ks of the X-ray observation
and obtained the simulated VHE fluxes for energies above 100 GeV, taking
advantage of the improved low-energy threshold of CTA with respect to MAGIC. We
note that the CTA performance simulation is for an observation at low zenith
angles, and a 4\,h observation is likely to cover a large range of zenith angles
resulting in an increased energy threshold. 

CTA is able to clearly
detect the fast variability exhibited by the X-ray light curve at scales down to
1000\,s, as well as recover the X-ray/VHE correlation, with $r\ge0.9$, for
flux variations of a factor 1.5. 

\begin{figure*}[]
    \includegraphics[width=0.50\linewidth,angle=0,clip]{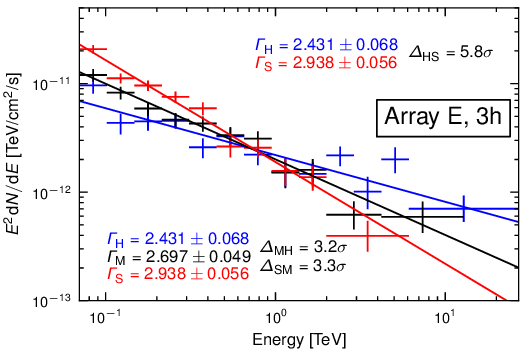}
    \includegraphics[width=0.50\linewidth,angle=0,clip]{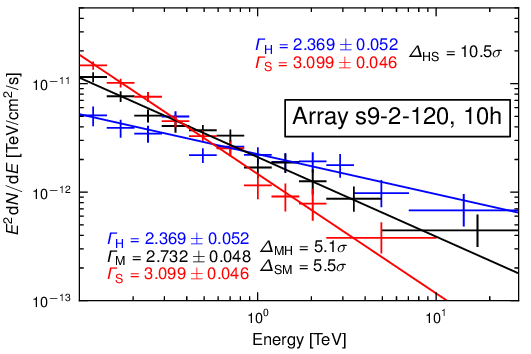}
    \caption{CTA simulations of  \lsi. \textit{Left:} Sensitivity of CTA
    \texttt{E} to spectral variations. Simulated spectra and power-law fits
    for 3\,hours of observation with the full CTA
    \texttt{E} array for each of the spectral states. The measured power-law
    indices and the significance of the variations are shown.  \textit{Right:} Sensitivity of CTA \texttt{s9-2-120} to spectral variations.  Simulated
    spectra and power-law fits for 10\,hours of observation with the subarray
    \texttt{s9-2-120} for each of the spectral states. The measured power-law
    indices and the significance of the variations are shown.
    \label{specsE}}
  \end{figure*}

The second observation scenario consists of a longer observational campaign of
about ten nights with three hours of observation per night, which we here
  simulate from the continuous 100\,ks X-ray light curve. Given the foreseeable pressure 
to obtain observation time with CTA, this campaign
would not be done with the full array but with a so-called subarray, a subset of
the 100 telescopes that may operate independently. Here we chose a subarray
configuration similar to an expanded H.E.S.S.: nine medium sized telescopes
located at a distance of 120\,m from each other, known as \texttt{s9-2-120}.
This kind of configurations will be readily available after the construction of
CTA begins, and thus the science case described could be achieved even before
the whole array becomes operational. In  Figure~\ref{corrsub} (left) we see that
the longer observation time allows the campaign to probe a wider range of X-ray
fluxes than the variation of a factor 1.5 used for the shorter, full CTA array
observation discussed above.
%, as well as different scales of variability from kiloseconds up to a few hours. 
Even with the reduced sensitivity of the subarray as compared to the full
array, the correlation is well recovered (see Fig.~\ref{corrsub} right), partly
owing to the aforementioned wider X-ray flux coverage.

To further explore the shortest time scales in which CTA can resolve a flare, we
simulated a 20\,hours event whose flux variation follows a Gaussian distribution and assuming the best-fit
spectral shape reported by MAGIC for the Cygnus X-1 signal \citep{cygx1}. 20\,hours represent the total duration of the flare (i.e. the mean of the Gaussian distribution is at 10 hours from the start of the observation). 
With a
binning of 5 min for each data point, CTA could clearly resolve this assumed Gaussian-shaped flare. A 5 min integration would result in a detection with a significance of 7$\sigma$ at the assumed low state and 25$\sigma$ in high state,
whereas with the sensitivity of MAGIC it is only possible to detect the peak of
this flare. This is a clear example of the better sensitivity of CTA with
respect to the existing IACTs. Although the limited duty cycle of Cherenkov
telescopes prevents them to observe a source for 20 hours in a row, a realistic
exposure time of 5 hours would be enough for CTA to resolve parts of the flare
in bins of $\sim$10 min.

As a conclusion from this simulation exercise, we see that the full CTA array will be
a powerful tool to probe into the fast flux variability of gamma-ray binaries, which
could allow the characterization of the dynamical processes taking place in the
emitting plasma. In addition, we have shown that subarray operations must be encouraged since
they are able to provide very interesting results using only a fraction of the
full array. They will prove invaluable not only in observational campaigns where the
long-term variability is important, but also to monitor flaring sources such as
gamma-ray binaries and AGNs.

\subsection{Sensitivity to spectral shape variations}

There have been some hints of VHE spectral variability of \lsi\ along the orbit
\citep{jogler09}, but they were not significant enough to be claimed as real.
The proper characterization of the spectral variability would allow the
characterization of the radiation mechanism, and more generally, as mentioned in
Section~\ref{error}, the determination of the conditions in the high-energy
emitter. To explore the power of CTA to disentangle different spectral indices,
we have considered a variation of the power-law VHE spectrum of \lsi\ within the
statistical error obtained in the MAGIC campaign on September 2007 (See
\cite{2009ApJ...706L..27A}) and test the time required by CTA to significantly
detect it. The measured spectral index was $\Gamma=2.7\pm0.3$, so we took the
two extreme values (2.4 and 3.0) and obtained the simulated CTA spectra. 
We computed
  the CTA spectra above the energy thresholds of 70~GeV and 100~GeV for
  the full array configuration \texttt{E} and the subarray
  \texttt{s9-2-120}, respectively.
 We found that
an observation with the full CTA \texttt{E} array of 3\,hours for each of the
spectral states (i.e., one night per state) provides a set of spectra from which
the photon index variations can be clearly seen. An example of such a
realization can be seen in Fig.~\ref{specsE}, left. 
The difference between the hard
($\Gamma_\mathrm{H}=2.4$) and soft ($\Gamma_\mathrm{S}=3.0$) is detected at a
confidence level of more than 3$\sigma$ in 99\% of the realizations,
with a mean value of $(7.8 \pm 1.8)\sigma$. Considering the variation with the mean
spectral shape of $\Gamma_\mathrm{M}=2.7$, the significance of the variation
with respect to the hard and soft spectra is lower but still above 3$\sigma$ for
85\% and 74\% of the realizations, respectively. We note that an increase in
observation time to 5\,hours per spectrum leads to a detection rate above 95\% for
all the spectrum pairs (see Figure~\ref{lsispecs}). We also considered the possibility of longer
observations with the subarray \texttt{s9-2-120}. In this case, for an observation of 10\,hours for each of the spectral
states, the spectral variation between the soft and mean spectra is
still significant at a level higher than $3\sigma$ for 82\% of the
simulation realizations. A representative realization of three simulated spectra with the
subarray \texttt{s9-2-120} can be seen in Fig.~6, right.

To further test the CTA spectral capabilities, we have used the derived spectrum
of Cygnus X-1 during the flare  to simulate 20 energy spectra with photon
indices ranging from $-$2 to $-$4.  We have also simulated different exposure
times: 5, 15, 30 and 60 min, to study the minimum
time scale to distinguish between the slope of different spectra.
Figure~\ref{sim_index_cygx1} left shows the  photon index error versus the simulated photon index in the fitting of the resulting CTA spectra.
CTA would be able to distinguish the different spectral slopes in
all cases except those showing the softest spectra, where the error bars
are too large to properly distinguish them at a high confidence level. In that
case, the observation of a flare as that reported in Cygnus X-1 would require
exposures $\geq$ 15 min.  Should such a kind of flare happen again, the minimum timescale for a 5
standard deviations detection of a flare within 10\% of the reported spectrum
from Cygnus X-1 is in the range of 2--3 min. However, in order to have a
spectrum determination, one can consider a 10$\sigma$ detection threshold. With
this constraint, the minimum timescale accessible is in the range of 8.5--12.5 min. This is shown in Fig~\ref{sim_index_cygx1} right.
The above estimates hold, provided that the responses of the array are as
stable as simulated for 30 minutes exposure and that the timescales are
probed a priori.

 \begin{figure}
  % \centering
  \includegraphics[width=1.0\linewidth,angle=0,clip]{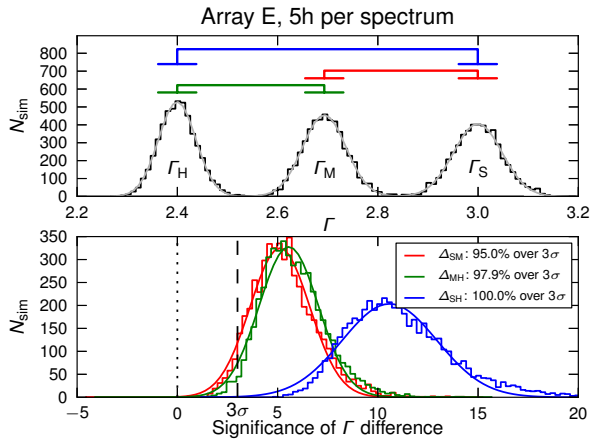}
 \caption{\label{lsispecs} Results of 10\,000 realizations of the spectral observation of LS~I~+61~303
with the array \texttt{E} during 5\,h per spectral state.\emph{Top:}
Distribution of recovered spectral indices for soft
($\Gamma_\mathrm{S}$), mean ($\Gamma_\mathrm{M}$), and hard
($\Gamma_\mathrm{H}$) spectra. \emph{Bottom:} Distribution of
significances of the difference between each of the spectrum pairs. The
legend indicates the fraction of realizations with a significance larger
than $3\sigma$.}
   \end{figure}

   \begin{figure*}
  % \centering
   \includegraphics[width=0.51\linewidth,angle=0,clip]{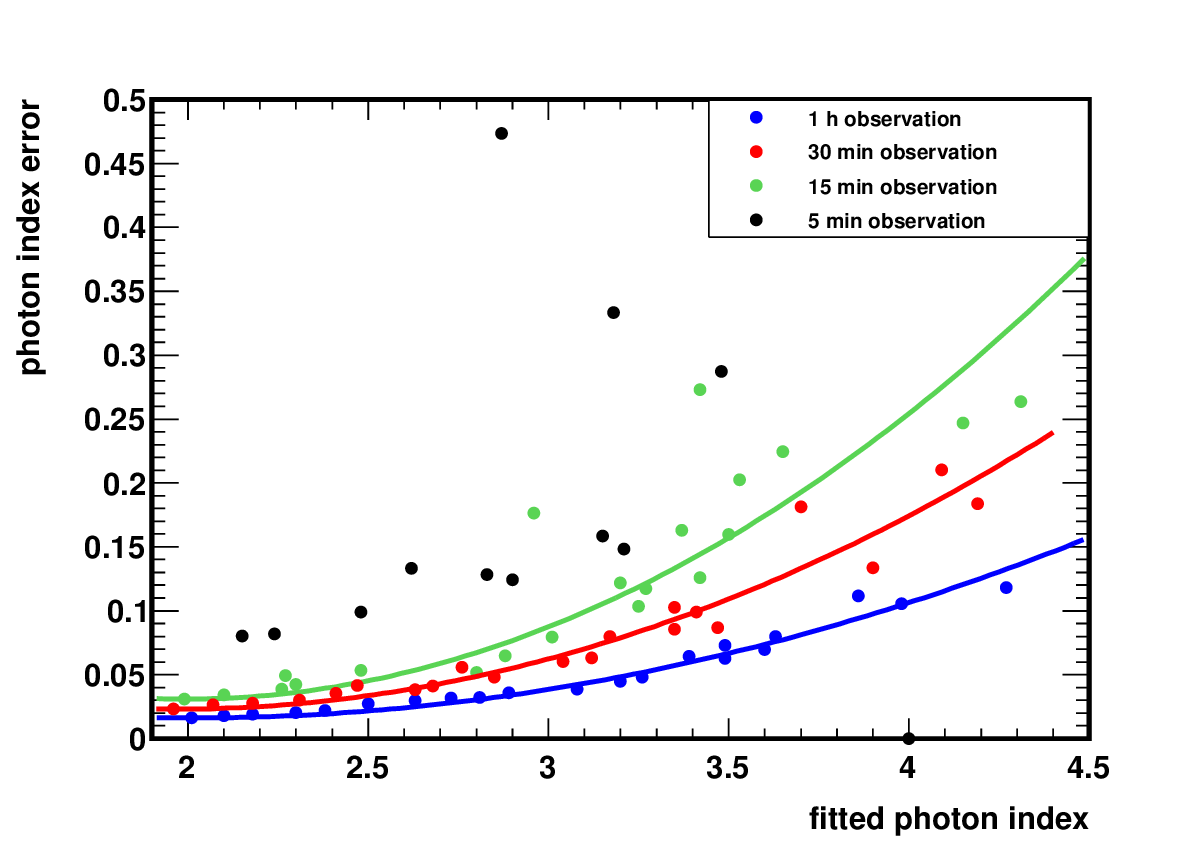}
   \includegraphics[width=0.48\linewidth,angle=0,clip]{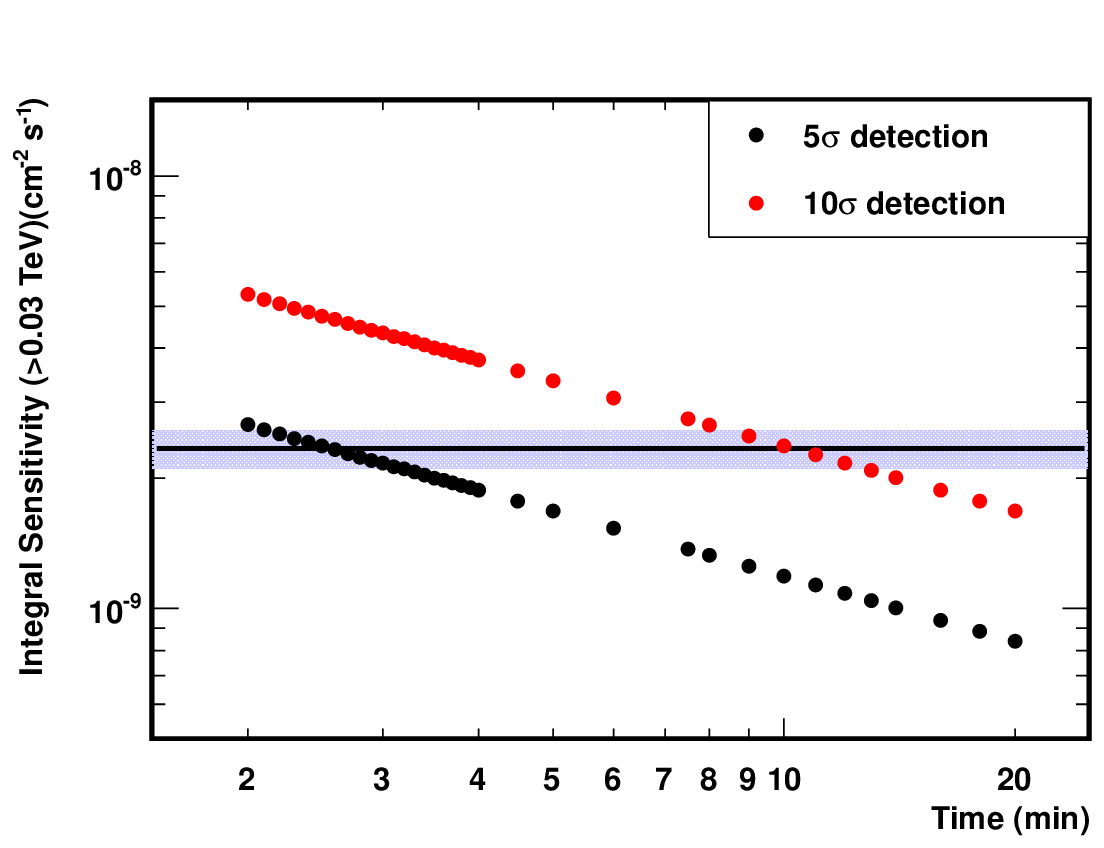}
     \caption{\label{sim_index_cygx1}CTA simulations of Cygnus X-1. \textit{Left: }Error on the photon index versus the fitted photon index. The photon index error can be understood as a proxy of the finest possible resolution.  The lines show the fit of a parabolic function to the data points normalized to the first point.  
      \textit{Right: }Integral sensitivity above 30\,GeV as a function of integration time for the
spectrum of Cygnus X-1 \citep{2007ApJ...665L..51A}, above 5 and 10 standard
deviations threshold (see text). The shaded area represents a variation of
10\% from the spectrum of Cygnus X-1 during the 2007 flare.}
    \end{figure*}

It is
clear that CTA will be a powerful tool for the detection of spectral variations
in gamma-ray binaries. The statistical errors we obtained for these simulations
are nearly an order of magnitude lower than the ones obtained with MAGIC in
2007, thus demonstrating the capability of CTA to deliver new and exciting
science in the following years.

\subsection{Exploring the minimum detectable time delay between X-ray
and TeV emission in gamma-ray binaries}

As stated in section~\ref{fast}, the MAGIC collaboration reported on the existence of a correlation between the X-ray emission and
VHE gamma-ray emission from the gamma-ray binary \lsi\ 
\citep{2009ApJ...706L..27A} (see however \citep{2011ApJÉ738É3A}).  The electron cooling times owing to synchrotron,
IC and adiabatic losses inside the system are expected to be around
a few thousand seconds (see, e.g. \citep{2011A&AÉ527AÉ9Z}). If there is
a time delay between the non-thermal emission at different bands (e.g., X-ray and
TeV) larger than the electron cooling time scale, the emission from these bands
would most likely have an origin in different locations in the binary. The
detection of such a delay has been impossible with the current generation of
IACTs, so here we present a study of the capability of CTA for such a study.
Furthermore, since CTA will operate together with a new generation of X-ray telescopes, such as the Japanese X-ray telescope {\it Astro-H}, we have
done this study considering the capabilities of this new instrument. LS I +61
303 is the binary system selected for this study.

 \begin{figure*}[]
   \centering
\includegraphics*[width=0.49\linewidth,angle=0,clip]{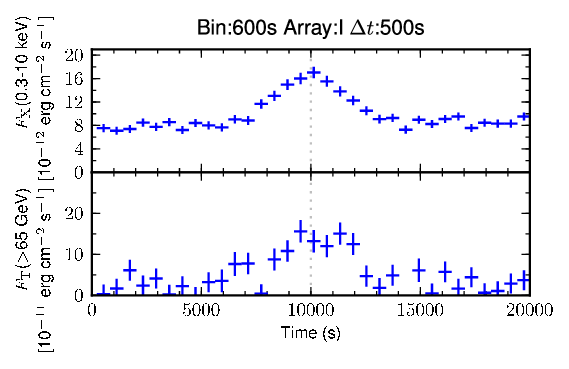}
\includegraphics*[width=0.49\linewidth,angle=0,clip]{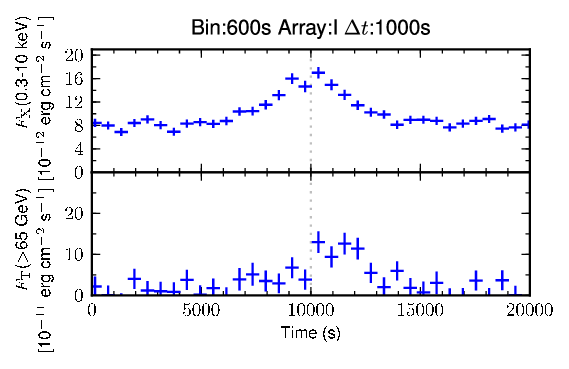}
\includegraphics*[width=0.49\linewidth,angle=0,clip]{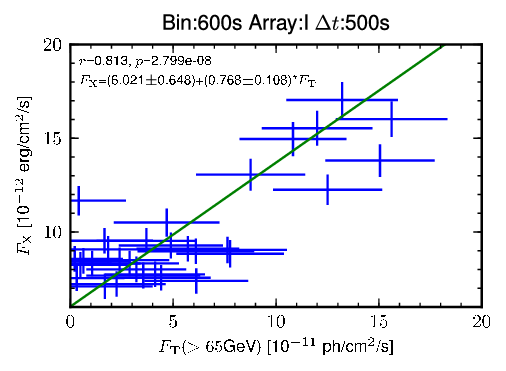}
  \includegraphics*[width=0.49\linewidth,angle=0,clip]{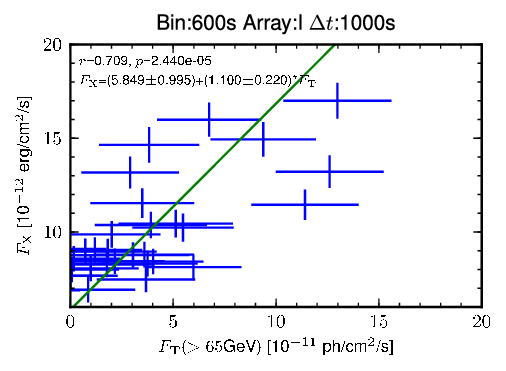}
 \includegraphics*[width=0.49\linewidth,angle=0,clip]{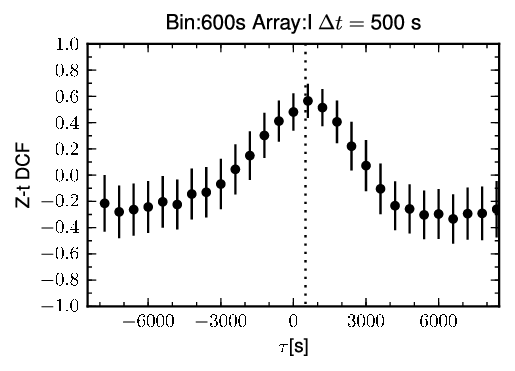}
 \includegraphics*[width=0.49\linewidth,angle=0,clip]{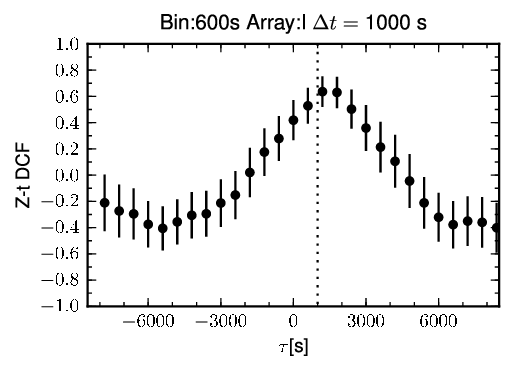}
   \caption{\label{delay}CTA simulations of  \lsi. \textit{Top left:}X-ray and VHE light curves of a Gaussian flare with sigma of 1500 s and 500-s delay. \textit{Top right:} same with 1000-s delay. \textit{Middle left:} X-ray flux vs.\ VHE flux for 500-s delay.  \textit{Middle right:} same with 1000-s delay. \textit{Bottom left:} ZDCF for 500-s delay. \textit{Bottom right:} same with 1000-s delay.}
    \end{figure*}
    
Based on the short X-ray flares from \lsi\ detected in
\citep{2010MNRAS.405.2206R}, we have chosen to model a Gaussian flare with a
width of 1500\,s. In the {\it Chandra} observations, these flares have a baseline
count rate of 0.35\,s$^{-1}$ and a peak of 0.8\,s$^{-1}$, corresponding to
unabsorbed energy fluxes of $\sim8$ and $\sim19$ times
$10^{-12}\,\mathrm{erg\,cm^{-2}\,s^{-1}}$, respectively. We used these parameters
to simulate the light curve detected by \emph{Astro-H}. 
Using the correlation between X-rays and VHE gamma rays found in
\citep{2009ApJ...706L..27A} and the tools for simulating the CTA response, we
generate the corresponding VHE light curves of the flare as seen by CTA  array above 65 GeV in configuration \texttt{I}.  
We have used a time binning of 600 s for both the CTA and the {\it Astro-H} light curves (LC), and studied positive delays
of the TeV light curve with respect to the X-ray light curve in the range 0 to 2000 s in steps of 100 s. We show in the top
panels of Figure~\ref{delay} the simulated light curves corresponding to 500-s delay (left) and 1000-s delay (right). In the
middle panels of Figure~\ref{delay} we show the X-ray fluxes as a function of the TeV fluxes in both cases. A low correlation coefficient is obtained due to a loop structure induced by the delay. When the delay is 0 the average correlation coefficient is r=0.86$\pm$0.04.

To clearly detect such delayed correlations, we have used the $z$-transformed
discrete correlation function (ZDCF), which determines 68\% confidence level intervals for the correlation coefficient for
running values of the delay (see, e.g. \cite{edelson88, alexander97}). We show in the bottom panels of Figure~\ref{delay}
the ZDCF in the case of 500-s delay (left) and 1000-s delay (right). The detection of correlated signals is clear in both cases. Since the VHE LC is generated from
the X-ray one, their errors are correlated and the scattering of one original (X-ray) light curve affects the second one
(VHE). To correct this problem we went through simulations: we started with 100 \emph{Astro-H} LC. With each of this 100 LC we produced 10 other X-ray LC by adding a Gaussian noise to the original ones, for each simulated delay. Then we simulated the corresponding VHE LC. At the end we had, for each delay, a sample of 1000 pairs of LC. For each pair of light curves we have calculated the ZDCF. To evaluate how significant is the measurement of a delay using the ZDCF we have fitted Gaussian functions to the maxima of the ZDCF. At the end we have 1000 values of the peak for each simulated delay distributed around the real delay (see Figure~\ref{histo}). The measured delay is then calculated as the mean value of the distribution and its uncertainty is the standard deviation. In Figure~\ref{sigdelay} we show the measured delay as a function of the simulated delay following the procedure described above. Considering all possible uncertainties, delays of $\sim$1000 s can be significantly detected at a 3$\sigma$ confidence level in simultaneous light curves obtained with  \emph{Astro-H} and CTA. These results are, to first order, independent of the duration of the short flares, as far as they last longer than the binning. Overall, these results indicate that CTA will allow us to localize and constrain the X-ray ant TeV emitting regions of gamma-ray binaries and their properties. 

%To evaluate how significant is the measurement of a delay using the ZDCF we have fitted Gaussian functions to the maxima of the ZDCF. 
%To estimate the delay and its uncertainty in each simulation, we have considered the delay value of the maximum of the Gaussian function and the standard deviation of the Gaussian function. 
%In Figure~\ref{sigdelay} we show the significance of the measured delay (delay divided by its uncertainty) as a function of the measured delay following the procedure described above. It is clear that delays as short as 500 s can be significantly detected in simultaneous light curves obtained with {\it Astro-H} and CTA. 
%Since this delay coincides with the original binning of the light curves, we have repeated these simulations using a binning of 250 s, and found that a significant correlation can still be found between the fluxes at X-ray and TeV energies, while delays as short as {\bf 400} s can be measured in a significant way. 
%The offset is due to the scatter of the X-ray light curve. Considering all possible uncertainties It is clear that delays as short as 400 s can be significantly detected in simultaneous light curves obtained with {\it Astro-H} and CTA. 

    \begin{figure}[]
\includegraphics*[width=1.0\linewidth,angle=0,clip]{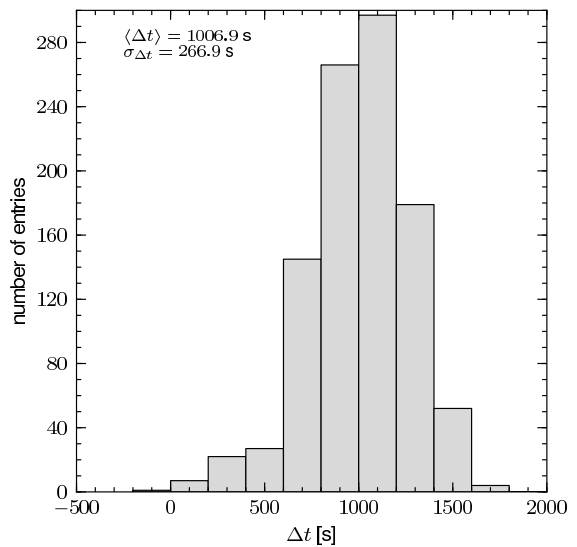}
   \caption{\label{histo} Distribution of the fitted peak of the ZDCF for the simulated light curves for an introduced delay of 1000 s.
   %(1000s, hi ha 2 figures). 
   A Gaussian fit gives the measured delay $\left<\Delta t\right>$ and its standard deviation $\sigma_{\Delta t}$.}
    \end{figure}

    \begin{figure}[]
\includegraphics*[width=1.0\linewidth,angle=0,clip]{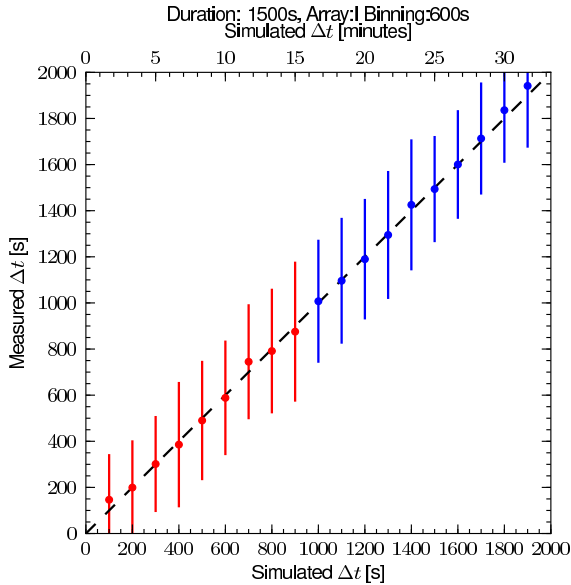}
   \caption{\label{sigdelay}CTA simulations of  \lsi. Measured time delay as a function of the simulated delay. Delays as short as  1000 s can be significantly measured.}
    \end{figure}

\subsection{Exploring the collision of microquasar jets with the interstellar medium}

 To simulate the CTA response to the observation of MQ jet/ISM interactions, we
 have used the theoretical predictions for a source with a jet power of $
 10^{38}$~erg~s$^{-1}$, a source age of $10^{5}$~yr embedded in a medium with
 particle density of $1$~cm$^{-3}$ \citep{2009A&A...497..325B}. Gamma-ray
 spectra with $\Gamma_{\rm ph} \sim 2.45$ and $\Gamma_{\rm ph} \sim 2.85$ are
 derived from current theoretical models for the leptonic and hadronic contribution,
 respectively \cite{2009A&A...497..325B, 2011MNRAS.410..978Z}. We studied the CTA performance in 50~h of observation time using
 the simulation tools for the \texttt{B} and \texttt{E} array configurations.
 Figure~\ref{finalspectrum} shows the obtained SED. The simulated flux is at
 a level $\sim1$\% of that of the Crab Nebula, although the steepening of the
 spectrum at high energies would make it difficult to detect the sources above a few TeV. 
 
 Regarding the extension of the emission as seen in gamma rays, accelerated
particles emitting at VHE do not have time to propagate to large distances since
radiative cooling is very effective. The emission will be mostly confined to the
accelerator region itself. The emitter size may not largely exceed the width of
the jet, $\sim 1$~pc, in the reverse shock region. In the case of the bow shock,
although it may extend sideways for much larger distances, only the region
around its apex ($\sim$~few~pc) will effectively accelerate particles up to the
highest energies. The total angular size of the emission from a source located
3~kpc away may then be $\lesssim$~few arcmin, and CTA would image a point-like
source with only a marginal extension roughly perpendicular to the jet
direction. 

To resolve the TeV emission produced in the jet/medium interaction regions and disentangle it from the putative
contribution produced close to the jet base, the reverse/bow shocks need to be located at a distance $\gtrsim$
$10^{19}$~cm. Although the precise location of the interaction regions
for a particular source can be difficult to predict since they depend on the jet power and age and the medium density
surrounding the system, hot spots displaying non-thermal emission at similar distances have been found, e.g. in SS~433
\cite{safi97}. Upper limits to the TeV emission from jet/medium interaction regions have been already reported for some microquasars, e.g. Cygnus X-1 \citep{cygx1} and GRS 1915+105 \citep{2009A&A...508.1135H}. However, clear evidences of non-thermal emission at the interaction sites in those cases are still lacking (see, e.g. \cite{2004ApJ...612..332K, 2005MNRAS.360..825Z, 2005Natur.436..819G}). 

Steady gamma-ray fluxes are predicted in  
jet termination regions. A constant flux pedestal level
could be observed on top of the orbital modulations expected from the central parts of the system for a wide range of orbital
geometries and system inclinations due to IC and pair-creation angle dependencies in their respective
cross-sections, and/or due to orbital variations of the accretion rate in both hadronic and leptonic scenarios. We note that CTA
may not be able 
to separate the possible contribution coming from the reverse shock and the forward (bow)
shocks. The shocked jet and medium material are separated by the contact discontinuity, and although the extent and relative
position of it with respect to both shocks is difficult to predict (see e.g. \cite{kaiser97}), it may be too short for the
angular resolution at the level of few arcminutes expected for CTA. Possible fast 
diffusion of accelerated particles behind the shocks
would complicate further the situation.

%\begin{center}
\begin{figure}[t]
\includegraphics[width=0.47\textwidth]{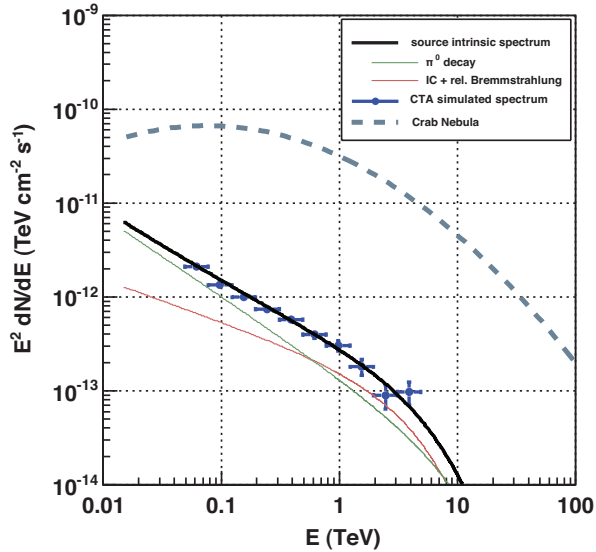}
%\begin{figure}[]
%\centering\includegraphics[width=6cm, angle=0]{figs/final_spectrum.eps}
\caption{CTA simulations of the emission from the MQ jet/ISM interaction regions. A jet power of $10^{38}$~erg~s$^{-1}$, a source age of $10^5$~yr and a medium with a particle density of~1~cm$^{-3}$ have been used. The gamma-ray spectra of IC and relativistic
Bremsstrahlung and that of p-p interactions through $\pi^ {0}$-decay for the adopted 
parameter values are adapted from \citep{2009A&A...497..325B} and
\citep{2011MNRAS.410..978Z}, respectively. Only the results from the \texttt{B} array configuration are displayed, 
since they are very similar to those obtained with the \texttt{E} array.\label{finalspectrum}}
\end{figure}
%\end{center}

\subsection{Exploring the colliding winds of massive star binary systems}

We performed numerical simulations of the response of CTA for a CWB like Eta Carinae. We based our simulations on the
measurements of the energy spectrum of Eta Carinae (see top panel of Fig.~\ref{carinae}) by the {\em Fermi}/LAT
\citep{2012arXiv1203.4939R} and the upper limits derived by the H.E.S.S. Collaboration \citep{
2012arXiv1204.5690H}. The spectrum between 0.1 and 100\,GeV is best fit by a power law with an exponential cutoff plus
an additional power law at high energies. In the TeV range, Eta Carinae has not been detected. In Figure
\ref{etacarsimul} (left) we show the Fermi/LAT data points and the H.E.S.S. upper limits in gray. From these measurements, it
seems that there must be a cutoff in the spectrum at high energies. For our simulations we assume exponential cutoffs 
at $E={100, 150, 200}$ GeV and test how well CTA could detect those. We produced simulations at increasing observation times
in order to study the minimal time required to detect the source and to get a meaningful spectra with such CTA observations.
In Figure~\ref{etacarsimul} (left), we show the simulated energy spectra with different cutoffs as they would be measured by
CTA. Simulations for 10 hours of observation time are displayed. To detect Eta Carinae, CTA would need 2--10 hours of
observations, depending on the energy cutoff in the spectrum; together with Fermi/LAT data, it should be possible to
determine the cutoff energy using a combined fit. However, it would take a longer time to determine the cutoff energy using
CTA data alone. The minimum observation time needed to significantly determine the cutoff energy, i.e.\ to distinguish
between a simple power law and a cutoff power law, is established using the likelihood ratio test for the two hypothesis. In
Figure~\ref{etacarsimul} (right), we show the resulting significance that a cutoff power law is a better fit to the data than
a pure power law versus integration time for the different energies of the cutoff. For this study we simulated 100 spectra
for each cutoff energy and for different observation times as shown in the plot. Taking 3$\sigma$ as a limit to distinguish
between the two different spectral hypothesis, one can see that 20\,hours are enough to detect the cutoff only if it is above 150
GeV. For a cutoff $\le$150\,GeV, 30 to 50 hours are needed. From our simulations, we can conclude that CTA observation times
of $>$15\,hours are necessary to make meaningful physics interpretation and modeling, whereas $>$20 hours are necessary
to precisely measure the energy cutoff in the spectrum. A proper characterization of the highest energy
cutoff will give important clues on the acceleration efficiency of the source, which may be operating close to the limit
predicted by diffusive shock acceleration, and on the nature of the radiation mechanism, either leptonic (IC) or
hadronic (proton-proton interactions). It is noteworthy that other colliding wind binary systems hosting powerful WR and O
stars may be also powerful non-thermal emitters, as hinted by hard X-ray observations or WR~140 with Suzaku \citep{2011BSRSL..80..724S}.

   \begin{figure*}
   \centering
   \includegraphics*[width=0.49\textwidth,angle=0,clip]{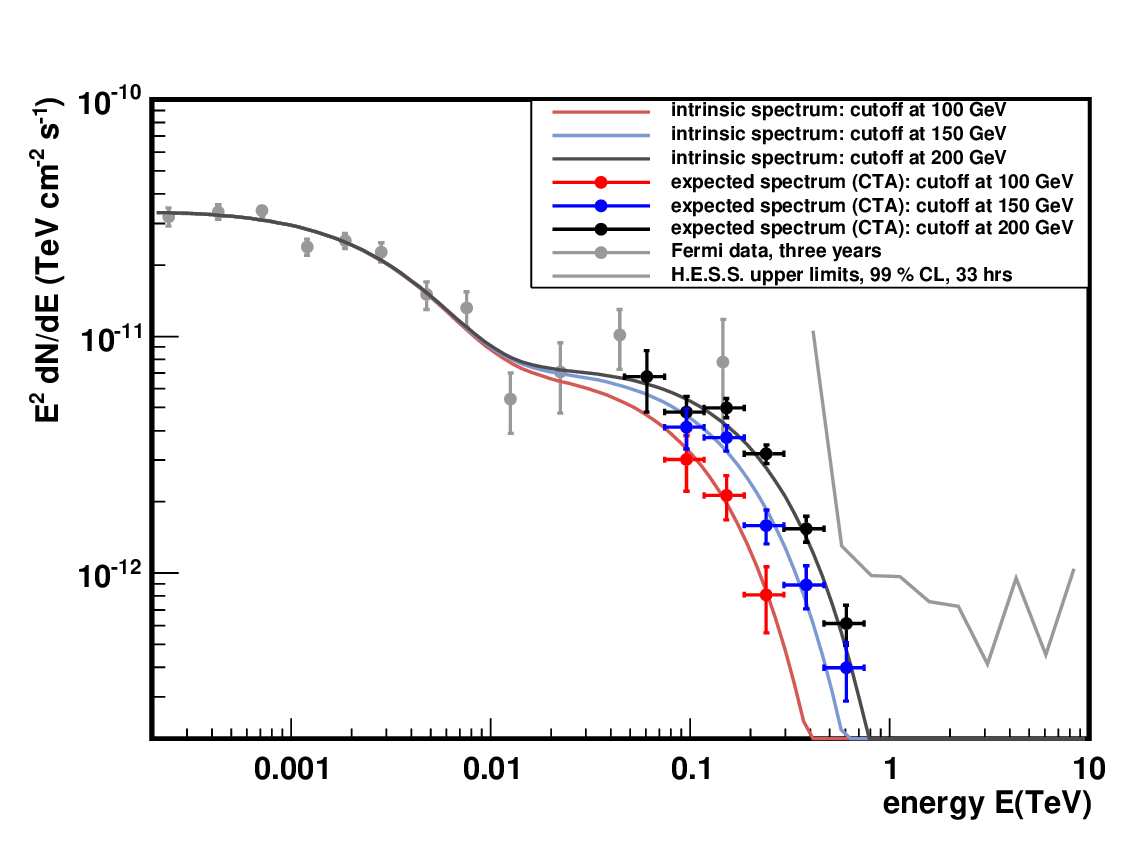}
   \includegraphics*[width=0.49\textwidth,angle=0,clip]{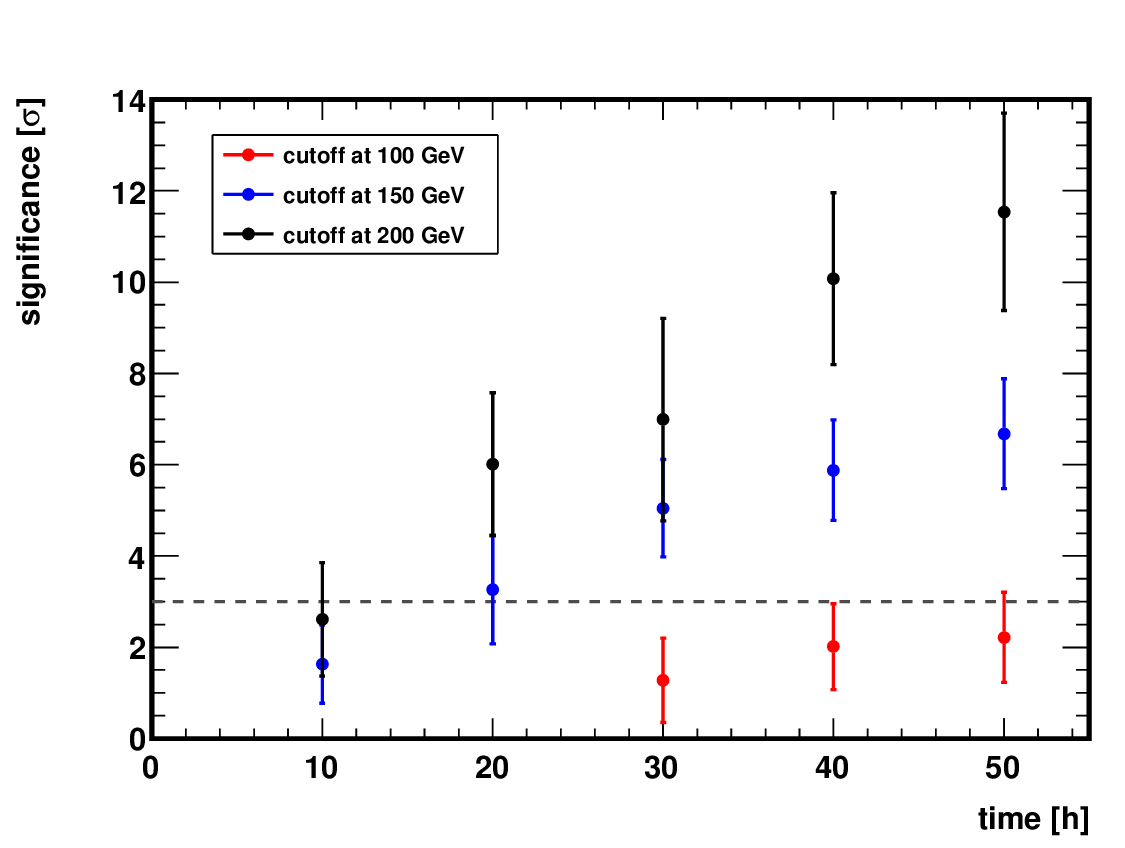}
   \caption{\label{fermi_cta_eta} CTA simulations of Eta Carinae. \textit{Left: }Intrinsic and CTA simulated spectra after 10 hours of observation time with high energy cutoffs at 100, 150 and 200\,GeV. \textit{Right: }Significance at which the corresponding cutoffs (color coded, see legend) can be resolved (i. e. are a better fit when compared to a simple power law) versus observation time. The dashed line at 3$\sigma$ is to guide the eye.}
   \label{etacarsimul}
    \end{figure*}

\section{Summary and conclusion}\label{summary}

The sensitivity of CTA   will lead to a very good sampling of light curves and spectra on very short timescales. It will
allow as well long source monitoring using subarrays, still with a sensitivity 2--3 times better than any previous instrument
operating at VHE energies. In particular, it is noteworthy that CTA will reduce by a factor of a few the errors in the
determination of fluxes and spectral indexes. The high sensitivity and good angular resolution will allow also for imaging of
possible extended emission in gamma-ray binaries, expected at the termination of the generated outflows. The low energy
threshold will also permit to study the maximum particle energy achievable in massive star binaries, trace the effects of
electromagnetic cascades in the spectra of gamma-ray binaries, or catch the most luminous part of the spectrum in some
sources. Finally, under CTA the population of gamma-ray binaries (and their different subclasses) may
easily grow by one order of magnitude, which will imply a strong improvement when looking for patterns and trends, tracing
the physical mechanisms behind the non-thermal activity in these sources. For all this, CTA, either in highly sensitive
observations of the whole array, or under the more suitable for monitoring subarray mode, will be a tool to obtain the
required phenomenological information for deep and accurate modeling of gamma-ray binaries.  This can mean a qualitative jump
in our physical knowledge of high-energy phenomena in the Galaxy.

~\

Acknowledgments: J.M.P., V.B-R, P.M.-A., J.M., M.R. and V.Z. acknowledge support by DGI of the Spanish Ministerio de Econom\'{\i}a y Competitividad (MINECO) under grants AYA2010-21782-C03-01 and FPA2010-22056-C06-02. J.M.P. acknowledges financial support from ICREA Academia. J.M. acknowledges support by MINECO under grant BES-2008-004564. M.R. acknowledges financial support from MINECO and European Social Funds through a Ram\'on y Cajal fellowship.  V.Z. was supported by the Spanish MEC through FPU grant AP2006-00077. P.B. has been supported by grant DLR 50 OG 0601 during
this work. V.B.-R. acknowledges the support of the European Community under a Marie Curie Intra-European fellowship and financial support from MINECO through a Ram\'on y Cajal fellowship. D.H., G.P and D.F.T. acknowledge support from the Ministry of Science and the Generalitat de Catalunya, through the grants AYA2009-07391 and SGR2009-811, as well as by ASPERA-EU through grant EUI-2009-04072. GD acknowledges support from the European Community via contract ERC-StG-200911. The research leading to these results has received funding from the
European Union's Seventh Framework Programme ([FP7/2007-2013]
[FP7/2007-2011]) under grant agreement n¡ 262053.



%% References


\bibliographystyle{model1-num-names}
\bibliography{app-cta}






% \end{thebibliography}

\end{document}